\documentclass[12pt]{article}

\usepackage{scicite}

\usepackage{times}

\newenvironment{sciabstract}{%
\begin{quote} \bf}
{\end{quote}}

\newcounter{lastnote}

\newcommand{\numberTotalNetworks}{163 }
\newcommand{\numberTotalGenes}{5112 }
\newcommand{\numberTotalConstants}{742 }

\newcommand{\numberNetworks}{122 }
\newcommand{\numberNetworksWithConstants}{94 }

\usepackage{xcolor}
\definecolor{Red}{rgb}{0.8,0,0}
\usepackage[normalem]{ulem}
\newcommand{\new}[2][]{\sout{#1}{\color{red}#2}}
\renewcommand{\new}[2][]{{#2}}

\usepackage{graphicx}
\usepackage{amsmath}
\usepackage{hyperref}
\usepackage{amssymb}

\newcommand{\beginsupplement}{%
        \setcounter{table}{0}
        \renewcommand{\thetable}{S\arabic{table}}%
        \setcounter{figure}{0}
        \renewcommand{\thefigure}{S\arabic{figure}}%
     }

\title{A meta-analysis of Boolean network models reveals design principles of gene regulatory networks}

\author
{Claus Kadelka$^{1\ast}$, Taras-Michael Butrie$^{2}$, Evan Hilton$^{3,4}$,\\ Jack Kinseth$^1$, Addison Schmidt$^{3}$, Haris Serdarevic$^{1}$\\
\\
\normalsize{$^{1}$Department of Mathematics, Iowa State University, Ames, IA 50011, United States,}\\
\normalsize{$^{2}$Department of Aerospace Engineering, Iowa State University, Ames, Iowa 50011, United States}\\
\normalsize{$^{3}$Department of Computer Science, Iowa State University, Ames, Iowa 50011, United States}\\
\normalsize{$^{4}$Bioinformatics and Computational Biology Program, Iowa State University, Ames, Iowa 50011, United States}\\
\\
\normalsize{$^\ast$To whom correspondence should be addressed; E-mail:  ckadelka@iastate.edu.}
}

\date{}

\begin{document} 


\baselineskip24pt


\maketitle


\begin{sciabstract}
Gene regulatory networks (GRNs) \new[control all]{play a central role in} cellular decision-making. Understanding their structure and how it impacts their dynamics constitutes thus a fundamental biological question. GRNs are frequently modeled as Boolean networks, which are intuitive, simple to describe, and can yield qualitative results even when data is sparse. We assembled the largest repository of expert-curated Boolean GRN models. A meta-analysis of this diverse set of models reveals several design principles. GRNs exhibit more canalization, redundancy and stable dynamics than expected. Moreover, they are enriched for certain recurring network motifs. This raises the important question why evolution favors these design mechanisms.
\end{sciabstract}



\section*{Introduction}

Gene regulatory networks (GRNs) describe how a collection of genes governs \new[the]{key} processes within a cell. Understanding how GRNs perform particular functions and do so consistently despite ubiquitous perturbations constitutes a fundamental biological question~\cite{stelling2004robustness}. Over the last two decades, a variety of design principles of GRNs have been proposed and studied, with a focus on discovering causal links between network form and function. 

GRNs have been shown to be enriched for certain sub-graphs with a specific structure, so-called network motifs, like feed-forward loops, feedback loops but also larger subcircuits~\cite{shen2002network,milo2002network,alon2007network,gerstein2012architecture}. Theoretical studies of the dynamic properties of these motifs revealed specific functionalities~\cite{mangan2003structure,cotterell2010atlas}. For example, coherent feed-forward loops can delay the activation or inhibition of a target gene, while incoherent ones can act as accelerators~\cite{mangan2003coherent}. Other hypothesized design principles include redundancy in the regulatory logic~\cite{nowak1997evolution,gu2003role}, and a high prevalence of canalization~\cite{kauffman1969metabolic, daniels2018criticality}. Canalization, a concept originating from the study of embryonal development~\cite{Wad}, refers to the ability of a GRN to maintain a stable phenotype despite ample genotypic as well as environmental variation.

Over the past decades, Boolean networks (reviewed in~\cite{schwab2020concepts}) have become an increasingly popular modeling framework for the study of biological systems, as they are intuitive and simple to describe. When data is sparse, as is still often the case for less-studied organisms and processes, complicated models (e.g., continuous differential equation models), which harbor the potential for quantitative predictions, cannot be appropriately fitted to the data due to their high number of parameters~\cite{karlebach2008modelling}. In this case, Boolean network models can often still yield qualitative results. 

Static network models are comprised of (i) a set of considered nodes (genes, external parameters, etc.), and (ii) a wiring diagram (also known as dependency graph), which describes which node regulates which and often also contains information about the respective type of regulation (positive due to e.g. transcriptional activation vs negative due to f.e. inhibition). A \emph{dynamic} Boolean network model possesses these same features but obtains its dynamics from an additional set of update rules (i.e., Boolean functions) that describe the regulatory logic governing the expression of each gene. Each gene is either on (i.e., high concentration, expressed) or off (i.e., low concentration, unexpressed) and time is discretized as well. 

Large, genome-wide \emph{static} transcriptional network models can be easily assembled from existing databases like TRANSFAC~\cite{matys2003transfac}, JASPAR~\cite{khan2018jaspar} or RegulonDB~\cite{gama2016regulondb}, by simply considering all known transcriptional regulations for a given species. However, information about the network topology alone provides only an incomplete understanding of a system, which is intrinsically dynamic. The formulation of dynamic models such as Boolean networks requires a careful calibration of the update rules by a subject expert. Therefore, all dynamic Boolean GRN models published thus far focus on specific biological processes of interest and contain only those genes involved in these processes~\cite{helikar2012cell}. \new{Moreover, most dynamic models have been published over the course of the last twelve years, as biological data needed for an accurate model description has become increasingly available. Over the course of the last few years, researchers have started to leverage the collection of these models to gain insights into specific aspects of GRNs such as the role of nonlinearity~\cite{manicka2023nonlinearity}, canalization~\cite{gates2021effective,subbaroyan2022minimum}, or the connection between canalization and criticality~\cite{daniels2018criticality, manicka2022effective, costa2023effective}.}

Here, we describe a \emph{comprehensive} meta-analysis of the largest repository of published, expert-curated Boolean GRN models assembled thus far. This provides a detailed understanding of the design principles of GRNs that are \new{potentially} conserved across organisms, and can help explain how GRNs operate smoothly and perform particular functions.

\section*{Results and discussion}\label{sec-results}

Using the biomedical literature search engine Pubmed, we created a database of \numberTotalNetworks Boolean GRN models. \new[We]{To avoid introducing bias into the meta-analysis, we} only included expert-curated models where the nodes and the update rules were selected by hand and not by a prediction algorithm or where default choices like threshold rules were used throughout. We further included only one version of highly similar models\new[, which]{This} led to the exclusion of 41 models (see Methods for details), resulting in a total of \numberNetworks models used in the meta-analysis, of which 61 are included in the Cell Collective~\cite{helikar2012cell} and 61 are not. The models describe \new[GRNs]{the regulatory logic underlying a variety of processes in numerous species} across multiple kingdoms of life (animals: 93, plants: 10, fungi: 9, bacteria: 9; Supplementary Dataset 1).

\begin{figure*} 
\centering
\includegraphics[width=7.25in]{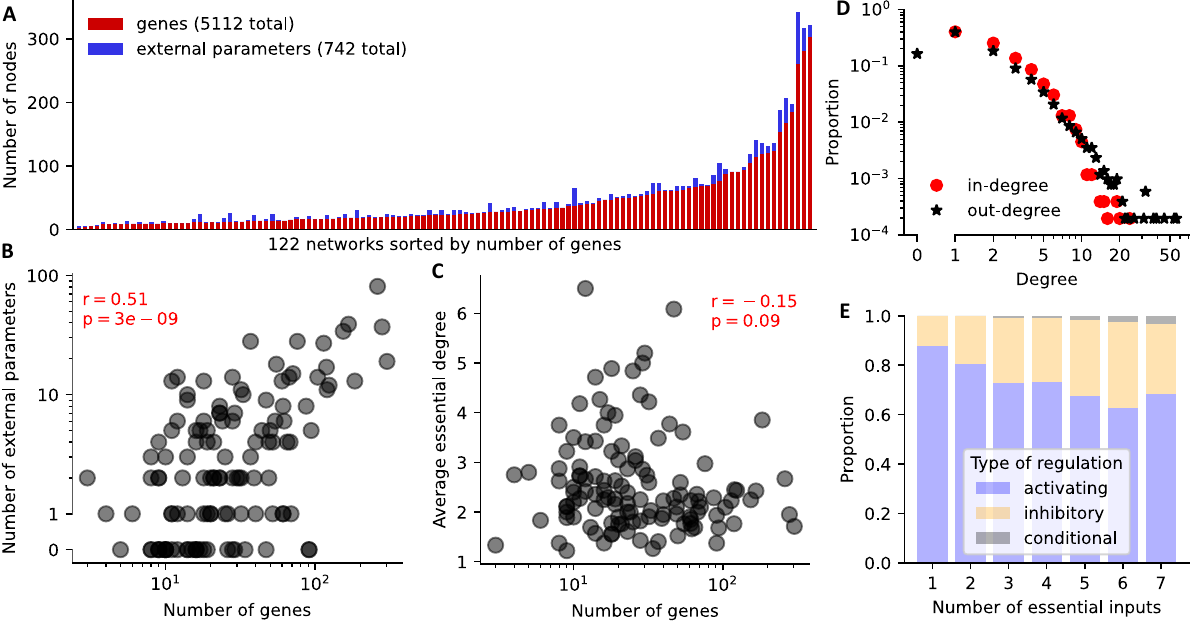}
\caption{{\bf Summary statistics of the analyzed GRN models}. (A) Plot of the number of genes and \new[constants (]{}external parameters\new[)]{} for each model sorted by number of genes. (B-C) For each model, the number of genes is plotted against (B) the number of \new[constants]{external parameters} and (C) the average essential in-degree of the genes. The Spearman correlation coefficient and associated p-value are shown in red. (D) In-degree (red circles) and out-degree (black stars) distribution derived from all \numberTotalGenes update rules. (E) Prevalence of each type of regulation (activation: blue, inhibition: orange, conditional: gray) stratified by the number of regulators (x-axis). \new{Non-essential regulations are excluded.}}
\label{fig:summary}
\end{figure*}

\new{The models contain different types of nodes. Some nodes are unregulated (i.e., they do not receive incoming edges in the wiring diagram) and remain thus constant over time. We refer to these nodes as \emph{external parameters} since they frequently represent abstract external conditions such as the temperature or pH level. Most other nodes represent genes. We therefore refer to all nodes that receive incoming edges in the wiring diagram as \emph{genes} but acknowledge that this is a simplification as some regulated nodes also represent molecules or abstract phenotypes such as cell proliferation or apoptosis.} The \new{\numberNetworks investigated GRN models} range\new{d} in size from $3$ to $302$ genes (mean = 41.9, median = 23), and encompass\new{ed} a total of 5112 genes as well as \numberTotalConstants external parameters\new[, which remain constant over time]{} (Fig.~\ref{fig:summary}A). \new{Some genes (as well as external parameters) appeared in multiple models (Supplementary Dataset 2), with \textit{AKT} appearing the most frequently, in 33 models.}

A majority of the investigated models (\numberNetworksWithConstants, $77\%$) contained external parameters\new[ such as temperature or pH levels, which are constant, i.e., not regulated]{}. As expected, network models with more genes contained on average more external parameters ($\rho_{\text{Spearman}} = 0.51$, Fig.~\ref{fig:summary}B). On the other hand, the size of a network was slightly negatively correlated with the average connectivity, i.e., the average number of regulators per gene ($\rho_{\text{Spearman}} = -0.15$, Fig.~\ref{fig:summary}C). The average connectivity differed widely across the \numberNetworks models; we observed a range of $[1.22,6.5]$ and a mean average connectivity of $2.56$ (median $=2.27$). The degree distribution of a random graph, in which the edges are distributed randomly, is a Poisson distribution~\cite{albert2002statistical}. When considering all update rules separately, we identified that the in-degree distribution resembled a Poisson distribution, while the out-degree distribution possessed a power-law tail (Fig.~\ref{fig:summary}D), as has been observed for many diverse types of networks~\cite{barabasi1999emergence,albert2002statistical}, including the yeast transcriptional regulatory network~\cite{guelzim2002topological}. The tails of the two degree distributions differed significantly; we found many more instances of high out-degree versus high in-degree, highlighting the presence of key transcription factors that act as network hubs~\cite{luscombe2004genomic,bemer2017cross}.

\begin{figure}[b!]
    \centering
    \includegraphics[width=\textwidth]{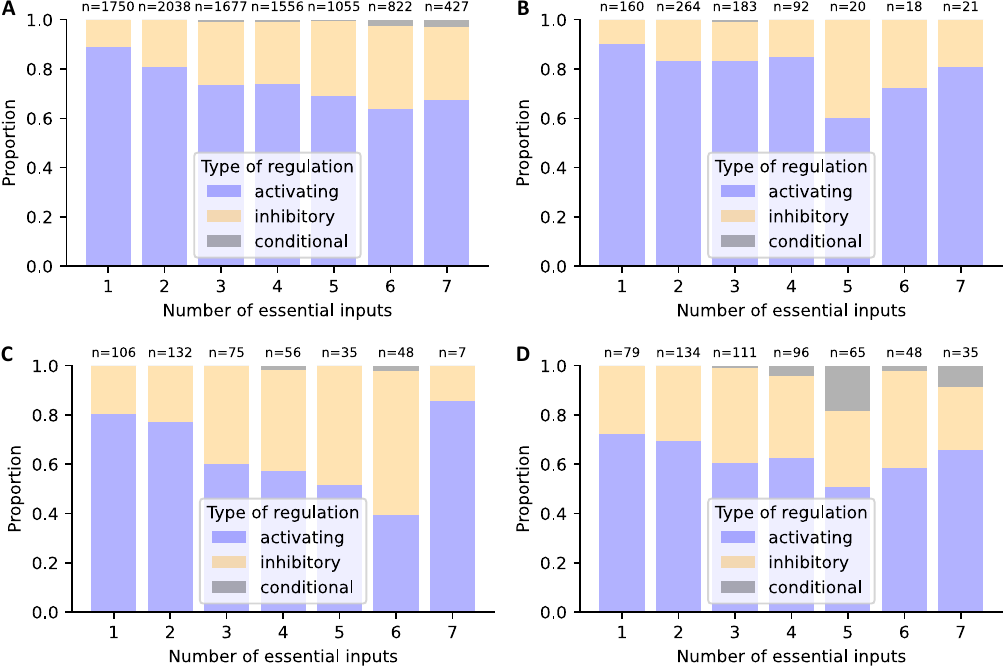}
    \caption{\new{{\bf Prevalence of different types of regulation per kingdom.} The prevalence of each type of regulation (activation: blue, inhibition: orange, conditional: gray) is shown. The analysis is restricted to rules with 1-7 essential inputs from Boolean GRN models of (A) animals, (B) bacteria, (C) fungi, (D) plants.
    The number of analyzed regulations is shown above each bar. The corresponding inter-kingdom analysis is shown in fig.~\ref{fig:summary}E.}}
    \label{fig:kingdoms_type}
\end{figure}

Next, we investigated the prevalence of different types of regulations. If gene A regulates gene B, there are three possibilities: (i) gene A may activate gene B, meaning that an increased expression of gene A (i.e., a change from $0$ to $1$ in the Boolean world) leads to an increased expression of gene B for some states of the other regulators, and \new[possibly to no]{possibly no} change in B for other states of the other regulators\new{,} (ii) gene A may inhibit gene B, meaning that an increase in A leads to a decrease in B for some states of the other regulators, and \new[possibly to no]{possibly no} change in B for other states of the other regulators, and (iii) gene A's effect on gene B may be conditional (i.e., not monotonic), meaning that for some states of the other regulators, A activates B, while for other states of the other regulators, A inhibits B. Except for two rules with more than 20 inputs, we investigated all update rules, resulting in a total of 12514 analyzed regulators \new{ (where some genes regulate more than one gene and each such regulation is considered separately)}.  
The majority of regulations were activations (9237, $73.8\%$), followed by inhibitions (2951, $23.6\%$) and conditional behavior (111, $0.9\%$). Regulatory networks in eukaryotes operate mainly by activation of otherwise inactive promoters~\cite{raeymaekers2002dynamics}. On the contrary, many promoters in prokaryotes are by default expressed and require repressors to reduce gene activity~\cite{struhl1999fundamentally}. Most of the considered GRN models are eukaryotic~(Supplementary Dataset 1), which \new[likely explains]{could serve as an explanation of} the increased prevalence of activation. \new{Surprisingly, we found, however, the largest proportion of activating interactions in bacterial (i.e., prokaryotic) GRN models ($656/785 = 83.6\%$), compared to $7844/10367 = 75.7\%$, $324/482 = 67.2\%$, ($401/640 = 62.7\%$ in GRNs of animals, fungi, and plants, respectively.} \new[We further found that activation seems]{Activation further seemed} particularly prevalent in situations where a gene's state is determined by one or only a few regulators (Fig.~\ref{fig:summary}E)\new[.]{, irrespective of the considered kingdom~(fig.~\ref{fig:kingdoms_type}).}

\begin{figure} 
    \centering
\includegraphics[width=0.6\textwidth]{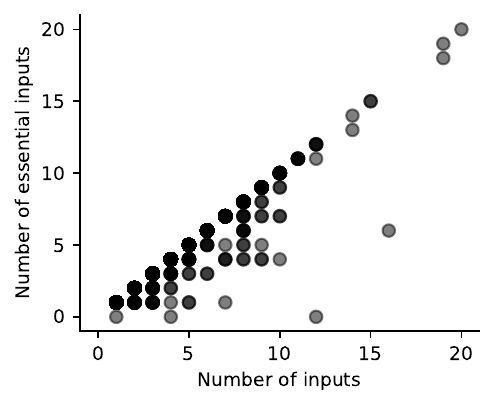}
\caption{{\bf Discrepancies in some published update rules}. For \numberTotalGenes update rules, the number of regulators in the identified published rule (x-axis) is plotted against the number of essential inputs after simplification of the rule (y-axis).}
\label{fig:mistakes}
\end{figure}

Interestingly, we found that 215 of the 12514 regulators ($1.7\%$) \new{contained in the ensemble of Boolean update rules} were non-essential\new{.} That is, these regulators \new[were part of]{appeared in} the published \new[update]{} rules but did not have any effect on the output. \new{For example, the Boolean update rule $(X\ AND\ Y)\ OR\ X$ simplifies to $X$; $Y$ is therefore a non-essential regulator.} \new[These]{The} non-essential regulators were spread across \new{23 ($18.9\%$) models and }120 ($2.3\%$) update rules\new[ and 23 ($18.9\%$) models]{, i.e., some update rules contained more than one non-essential regulator}. \new{In one extreme case, an update rule with twelve different inputs simplifies to the Boolean zero function.} \new{Fig.}{Figure}~\ref{fig:mistakes} shows the discrepancy between the number of inputs in the published update rules and the number of inputs that have an actual effect on the dynamics. \new[In one extreme case, an update rule with twelve different inputs simplifies to the Boolean zero function.]{}In the rest of this paper, only essential regulators were considered.

\begin{figure*} 
\centering
\includegraphics[width=5.67in]{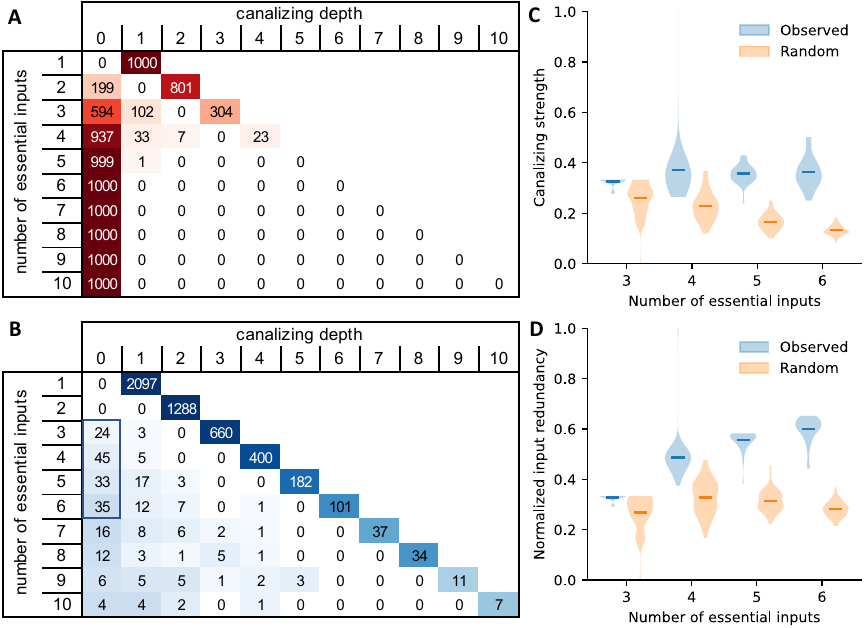}
\caption{{\bf High prevalence of canalization}. (A) Expected distribution of the canalizing depth for random Boolean functions for different numbers of essential inputs (1-10), based on 1000 random functions each. (B) Stratification of all identified update rules based on the number of essential inputs (rows) and the canalizing depth (columns). Update rules with more than ten inputs were omitted here, see table~\ref{tab:canalization_full} for a full table. The color gradient in A and B is computed separately for each row. (C-D) The distribution of the (C) canalizing strength and (D) normalized input redundancy of all observed 0-canalizing functions with 3-6 essential inputs (that is, all functions in the blue box in B) is shown (blue), as well as the expected distribution for random Boolean functions (orange), derived from 1000 samples each. Horizontal lines depict the respective mean values.}
\label{fig:canalization}
\end{figure*}

\subsection*{Canalization}
The concept of canalization, already introduced in the 1940s in the context of embryonal development~\cite{Wad}, has been proposed as a possible explanation for the remarkable stability of GRNs in the face of ubiquitous perturbations~\cite{gibson2000canalization,hallgrimsson2019developmental}. Accordingly, Boolean canalizing functions have been proposed as suitable update functions in Boolean GRN models~\cite{kauffman1974large}. Recently, the class of canalizing functions has been further stratified and studied~\cite{he2016stratification,kadelka2017influence}. Some smaller studies support the general hypothesis by revealing an overabundance of canalizing functions in GRN models~\cite{harris2002model,daniels2018criticality} but a rigorous, comprehensive analysis that considers various types of canalization is still missing.

A canalizing function possesses at least one input variable such that, if this variable takes on a certain ``canalizing" value, then the output value is already determined, regardless of the values of the remaining input variables. If this variable takes on another value, and there is a second variable with this same property, the function is $2$-canalizing. If $k$ variables follow this pattern, the function is $k$-canalizing~\cite{he2016stratification}, and the number of variables that follow this pattern is the canalizing depth of the function~\cite{layne2012nested}. If the canalizing depth equals the number of inputs (i.e., if all variables follow the described pattern), the function is also called a nested canalizing function\new[ (NCF)]{}. 

\begin{figure}
    \centering
    \includegraphics[width=\textwidth]{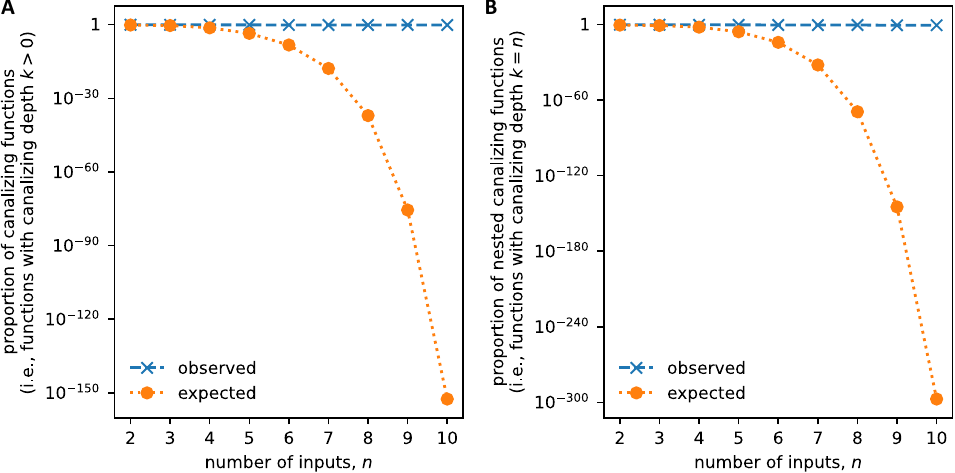}
    \caption{\new{{\bf Proportion of canalizing and nested canalizing functions.} For Boolean functions with $2-10$ inputs (x-axis), the proportion of (A) canalizing functions and (B) nested canalizing functions observed in published expert-curated GRN models (blue x) is compared to the expected proportion (orange dots), which is computed using explicit formulas for the number of canalizing and nested canalizing functions from~\cite{he2016stratification}.}}
    \label{fig:canalizing_proportion}
\end{figure}

To test the level of canalization in published GRN models, we stratified all \numberTotalGenes update rules based on their number of essential inputs and their canalizing depth. The number of Boolean functions with a certain canalizing depth is known~\cite{he2016stratification}, and the fraction of random Boolean functions which are canalizing (i.e., those with canalizing depth $\geq 1$) decreases \new[quickly]{exponentially} as the number of inputs increases (Fig.~\ref{fig:canalization}A). Most identified update rules, however, possessed a high canalizing depth, even rules with many inputs (Fig.~\ref{fig:canalization}B). 4827 out of the 5110 investigated update rules ($94.4\%$) were even nested canalizing, meaning that all their variables become ``eventually" canalizing~\cite{dimitrova2022revealing}. \new{A comparison of the expected and observed proportion of canalizing and nested canalizing functions reveals the true significance of the overabundance of canalization in GRN models~(fig.~\ref{fig:canalizing_proportion}).} These findings agree with earlier, smaller studies~\cite{harris2002model,daniels2018criticality}, which focused solely on the abundance of canalizing and nested canalizing functions but lacked the finer level of detail added by the canalizing depth. 

Our findings raised an important question: Are biological networks enriched for canalizing functions solely because of the strong overabundance of nested canalizing functions, or is there broader evidence for canalization in general\new[.]{?} To answer this, we relied on a broader mathematical definition of the biology-inspired concept of canalization, called collective canalization~\cite{Reichhardt}. Rather than focusing on single inputs that determine the output of a function regardless of the values of the remaining inputs, we studied the proportion of \emph{sets} of inputs that have this  \emph{canalizing} ability. The recently introduced canalizing strength of a Boolean function summarizes this information in a single measure~\cite{kadelka2023collectively}. By comparing the canalizing strength of all identified non-canalizing update rules with 3 to 6 inputs (i.e., those with canalizing depth $0$) with random non-canalizing Boolean functions, we found that even those update rules, non-canalizing according to Kauffman's stringent definition of canalization~\cite{kauffman1974large}, exhibited a higher level of collective canalization than expected~(Fig.~\ref{fig:canalization}C). Published non-canalizing update rules also exhibited more than expected input redundancy (Fig.~\ref{fig:canalization}D), which is an alternative measure of collective canalization~\cite{gates2021effective}.

\begin{figure} 
\centering
\includegraphics[width=3.55in]{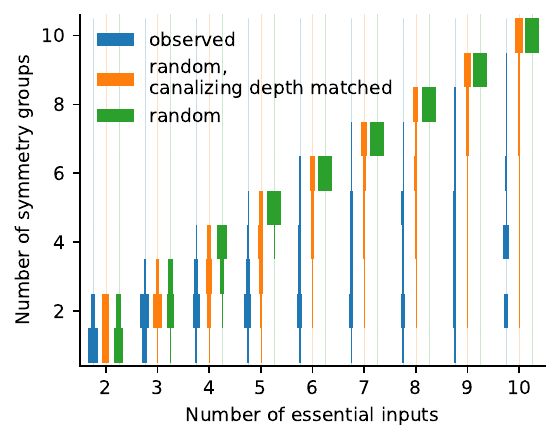}
\caption{{\bf High prevalence of redundancy}. The empirical distribution of the redundancy, measured by the number of symmetry groups (y-axis), is computed for all identified update rules (blue), stratified by the number of essential inputs (x-axis). For comparison, the expected distribution of the number of symmetry groups for random Boolean functions with 1-10 essential inputs is included (green), as well as the expected distribution for random Boolean functions with the same canalizing depth distribution as observed update rules (orange), as shown in Fig.~\ref{fig:canalization}A. Each expected distribution was generated using 1000 random functions. fig.~\ref{fig:redundancy_detail} contains the explicit values of each distribution.}
\label{fig:redundancy}
\end{figure}

\subsection*{Redundancy}
\begin{figure*} 
\centering
\includegraphics[width=0.99\textwidth]{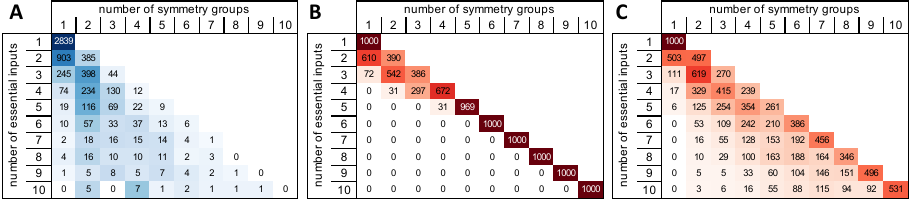}
\caption{{\bf Prevalence of redundancy}. (A) Stratification of all identified update rules based on the number of essential inputs (rows) and the redundancy, measured by the number of symmetry groups (columns). Update rules with more than ten essential inputs were omitted. (B-C) Expected distribution of the number of symmetry groups for random Boolean functions with 1-10 essential inputs. For each row, 1000 random, non-generated functions were generated. In (C), the distribution of the canalizing depth of the random, non-degenerated Boolean functions was matched to the one observed for the GRN models, shown in Fig.~\ref{fig:canalization}A.}
\label{fig:redundancy_detail}
\end{figure*}

Genetic redundancy constitutes an important feature of gene regulation, as the presence of duplicate genes provides robustness against null mutations~\cite{nowak1997evolution,gu2003role}. We tested the level of redundancy contained in the GRN models by quantifying the number of symmetry groups for each update rule. Two regulators are in the same symmetry group if they have exactly the same effect on the targeted gene, for all possible states of all other regulators. Redundant genes perform the same function and would thus be part of the same symmetry group. We found a much higher level of redundancy in the biological networks (i.e., much fewer symmetry groups; Fig.~\ref{fig:redundancy} and fig.~\ref{fig:redundancy_detail}A) than expected by chance (fig.~\ref{fig:redundancy_detail}B). This comparison is skewed since canalizing functions possess on average fewer symmetry groups. To exclude this confounding effect of canalization, we considered random functions whose canalizing depth was drawn from the empirical canalizing depth distribution of the published update rules~(fig.~\ref{fig:redundancy_detail}C). Even after this correction, published models exhibited a substantially higher level of redundancy (Fig.~\ref{fig:redundancy}).

\subsection*{Feed-forward loops}
\begin{figure} 
\centering
\includegraphics[width=7.25in]{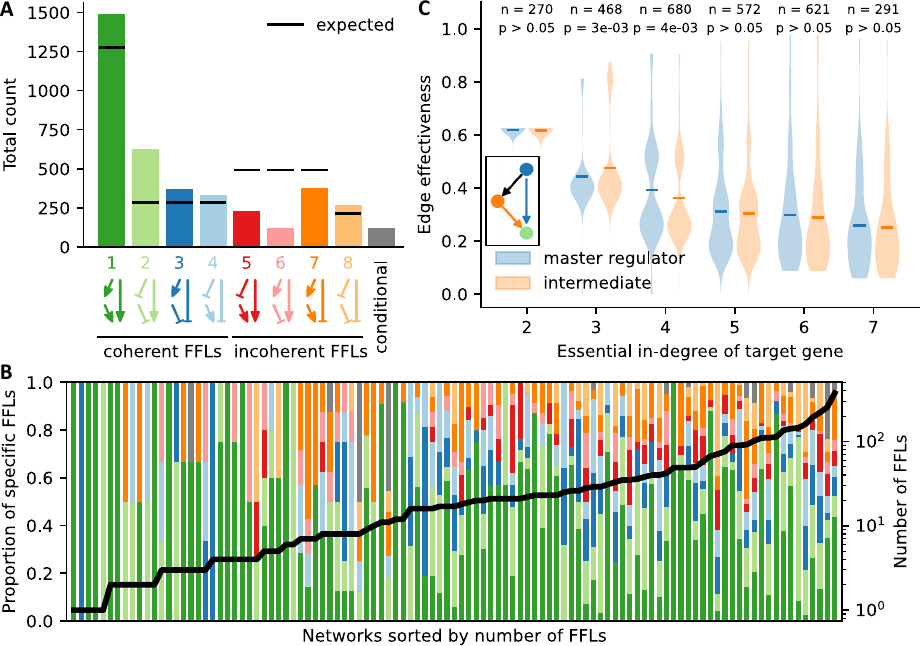}
\caption{{\bf Abundance of coherent feed-forward loops}. (A) Total number of the different types of FFLs in the \numberNetworks GRNs (colored bars). Conditional FFLs (gray) contain at least one conditional regulation preventing the determination of their exact type. Black horizontal lines indicate the respective expected number, which is based on null model 1 (see Methods). Type 1-4 FFLs are coherent, while type 5-8 FFLs are incoherent. (B) Proportion (stacked bar, color-coded as in A) and total number (black line) of the different types of FFLs for each network. The 17 networks without any FFLs are omitted. (C) For each target gene in a FFL (green), the edge effectiveness of the master regulator (blue) and the intermediate regulator (orange) is compared, stratified by the essential in-degree of the target gene. Horizontal lines depict the respective mean values. n = number of target genes with given essential in-degree, p = p-value from a two-tailed Wilcoxon signed-rank test.}\label{fig:ffl}
\end{figure}

Network motifs are sub-graphs with a specific structure that recur throughout a network and often carry out a certain function~\cite{milo2002network,alon2007network}. Several network motifs are commonly found in large, static GRN models such as the transcriptional network of \textit{E. coli}~\cite{shen2002network}. One such motif is the feed-forward loop (FFL), which consists of three genes: one master regulator that regulates both other genes, one target gene that is jointly regulated by both others, and one intermediate gene. In a \emph{coherent} FFL, the direct effect of the master regulator on the target has the same sign, either positive or negative, as the net indirect effect through the intermediate gene. Otherwise, the FFL is \emph{incoherent} (Fig.~\ref{fig:ffl}A displays all eight FFL types). Incoherent FFLs may act as sign-sensitive accelerators of the expression of the target gene, while coherent FFLs act as sign-sensitive delays~\cite{mangan2003structure}. Here, sign-sensitive means that the motif performs a function only in one direction, either when the target is up- or downregulated.

\begin{figure} 
    \centering
\includegraphics[width=0.49\textwidth]{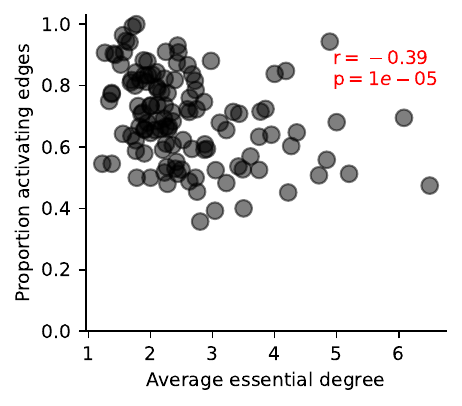}
\caption{{\bf Highly connected models feature a lower proportion of activating edges}.  For each model, its average essential in-degree is plotted against its proportion of activating edges (out of all activating and inhibitory edges, excluding conditional and non-essential edges). The Spearman correlation coefficient and associated p-value are shown in red.}
\label{fig:degree_vs_prop_pos}
\end{figure}

\begin{figure}
    \centering
    \includegraphics[width=\textwidth]{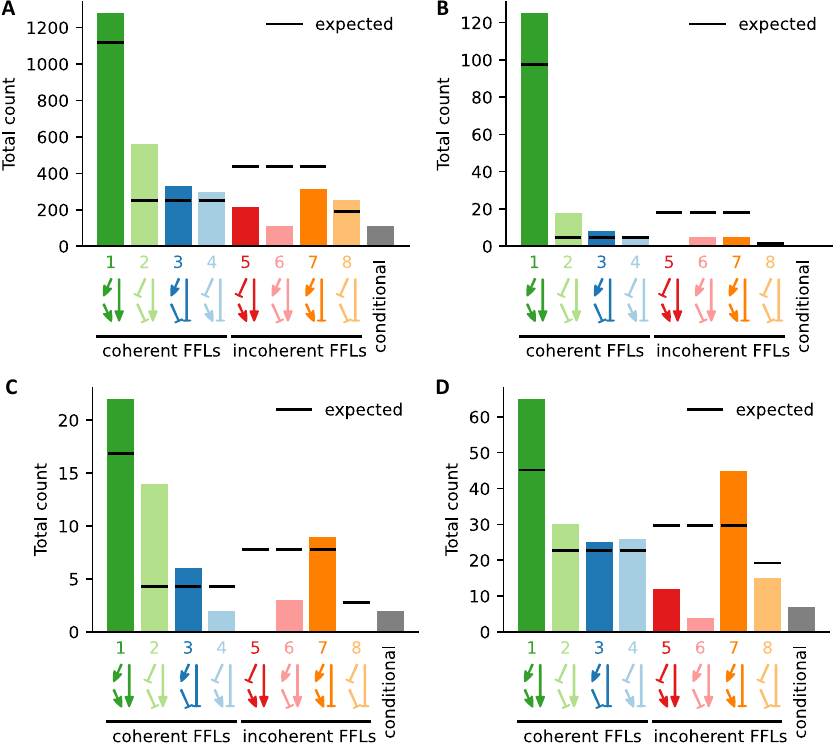}
    \caption{\new{{\bf Prevalence of feed-forward loops per kingdom.} The number of the different types of FFLs in GRN models of (A) animals, (B) bacteria, (C) fungi, (D) plants (colored bars) is shown. Conditional FFLs (gray) contain at least one conditional regulation preventing the determination of their exact type. Black horizontal lines indicate the respective expected number, which is based on null model 1 (see Methods). Type 1-4 FFLs are coherent, while type 5-8 FFLs are incoherent. The corresponding inter-kingdom analysis is shown in Fig.~\ref{fig:ffl}A.}}
    \label{fig:kingdoms_ffls}
\end{figure}

We identified a total of 3938 FFLs in the GRN models and stratified the number of occurrences by type~(Fig.~\ref{fig:ffl}A) and additionally by model~(Fig.~\ref{fig:ffl}B, \new{~Supplementary Dataset 3}). 122 FFLs ($3.1\%$) contained conditional regulations, which means that the type of these loops changes dynamically. The expected number of activating versus inhibitory regulations contained in a FFL depends on the proportion of activating regulations in the GRN models. This proportion varies strongly from model to model~(fig.~\ref{fig:degree_vs_prop_pos}) and decreases on average for models with higher degree. We therefore computed an expected number of each FFL type for each model, which we then summed up to obtain a total number (see Methods). 
Overall, the GRN models were enriched for each type of coherent FFL~(Fig.~\ref{fig:ffl}A). \new{This finding was consistent across kingdoms~(fig.~\ref{fig:kingdoms_ffls}).} All coherent FFL types, most involving two inhibitions, were almost as frequent as any incoherent FFL type, most of which contain only one inhibition.
Moreover, the incoherent FFL with three inhibitory regulations (type 8) was more prevalent than two of the three incoherent FFL types with only one inhibitory regulation. It was even the only incoherent FFL, which appeared more frequently than expected.

As reported for the static GRN models of \textit{E. Coli} and \textit{S. cerevisiae}~\cite{mangan2003structure} and as expected by chance, the FFL with three activating edges (type 1) proved by far the most prevalent. Interestingly, the type 2 FFL far outnumbered the remaining FFL types, including the two other coherent ones. This is surprising as coherent FFLs of types 2-4 all contain one activating and two inhibiting edges. The only potential explanation is that type 2 FFLs induce a positive effect on the target gene, while the effect is negative in type 3 and type 4 FFLs. Another interesting observation relates to type 6 FFLs. While these FFLs outnumbered all other incoherent FFLs (types 5,7,8) in the static GRN models of \textit{E. Coli} and \textit{S. cerevisiae}~\cite{mangan2003structure}, we found FFL type 6 to be the least abundant. This may be due to low sample sizes in the earlier publication, or due to genuine differences in genome-wide transcriptional networks versus dynamic GRN models, which focus on a relatively small subset of genes involved in a certain biological process of interest. To explain all these observed differences, theoretical studies similar to~\cite{mangan2003coherent,kaplan2008incoherent} may be needed, which focus on the functions of the different types of FFLs in dynamic GRN models.

The target gene in a FFL is regulated by both the master regulator and the intermediate regulator. To test if one of these two regulations is generally more important, we compared their edge effectiveness, which captures the extent to which a given input (i.e., an edge) is on average necessary to determine the value of a Boolean function~\cite{gates2021effective}; an important input has high edge effectiveness. As inputs to functions with more variables generally have lower edge effectiveness, we stratified the analysis by the essential in-degree $k$ of the target gene. Albeit weakly significant but opposing differences for $k=3$ (two-tailed Wilcoxon signed-rank test; p=0.003) and $k=4$ (p=0.004), we did not find any support for the hypothesis that either the master regulator or the intermediate regulator in a FFL is generally more important.

\begin{figure} 
\centering
\includegraphics[width=5.67in]{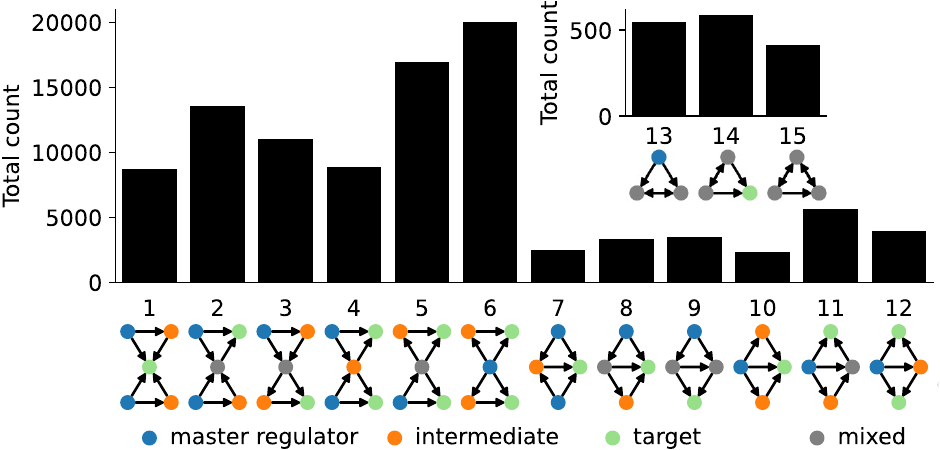}
\caption{{\bf Abundance of clusters of feed-forward loops}. Total number of the different types of clusters of FFLs in the \numberNetworks GRN models. Nodes in the motif graphs are color-coded based on their role in the two clustered FFLs: master regulators (blue), intermediate genes (orange), target genes (green), genes that appear in both FFLs but with a different role (gray).}
\label{fig:ffl_ffl}
\end{figure}

We further investigated the occurrence of clusters of FFLs, that is, two FFLs that share at least one node. As with single FFLs, we can distinguish different types of FFL clusters based on the distribution of activating and inhibiting edges in the motif (Fig.~\ref{fig:ffl_ffl} displays all 15 types of FFL clusters). A recent analysis of a diverse set of natural and engineered networks revealed wide differences in the distribution of the different types of FFL clusters~\cite{gorochowski2018organization}. 

\begin{figure*} 
\centering
\includegraphics[width=0.99\textwidth]{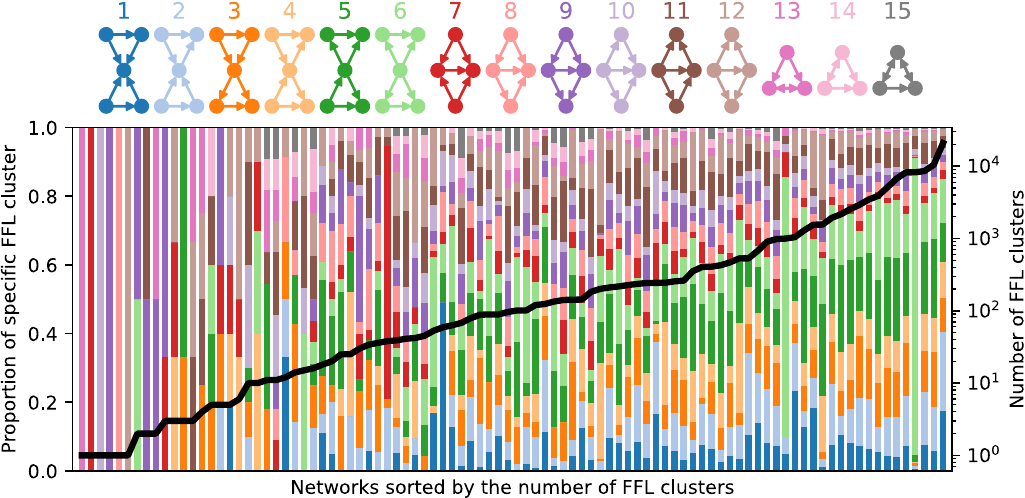}
\caption{{\bf Abundance of clusters of feed-forward loops per model}. Proportion (color-coded stacked bar) and total number (black line) of the different types of FFL clusters for each network. The 34 networks without any FFL clusters are omitted.}
\label{fig:ffl_ffl_detail}
\end{figure*}

We identified a total of 101832 FFL clusters in the \numberNetworks GRN models\new{~(Supplementary Dataset 4)}. As with the single FFL motifs, we stratified the number of occurrences by type~(Fig.~\ref{fig:ffl_ffl}) and additionally by model~(fig.~\ref{fig:ffl_ffl_detail}). As expected, we found most FFL clusters to involve 5 genes (79115, $77.7\%$), followed by 4 (21168, $20.8\%$) and by 3 genes (1549, $1.5\%$). As in the transcriptional networks of \textit{E. coli} and \textit{S. cerevisiae}~\cite{gorochowski2018organization}, type 6 was the most abundant. This type of FFL cluster features a master regulator involved in both FFLs and its abundance is likely due to the known presence of transcription factor hubs, which was also observed in this meta-analysis (Fig.~\ref{fig:summary}D). Type 11 was the most abundant among all FFL clusters involving 4 genes. This is surprising since transcriptional networks of \textit{E. coli} and \textit{S. cerevisiae} contained almost exclusively type 12 and hardly any of the other 4-gene FFL clusters~\cite{gorochowski2018organization}. An explanation for these discrepancies likely requires novel theoretical or computational studies that relate motif structure to motif function.
 
\subsection*{Feedback loops}
\begin{figure} 
\centering
\includegraphics[width=0.99\textwidth]{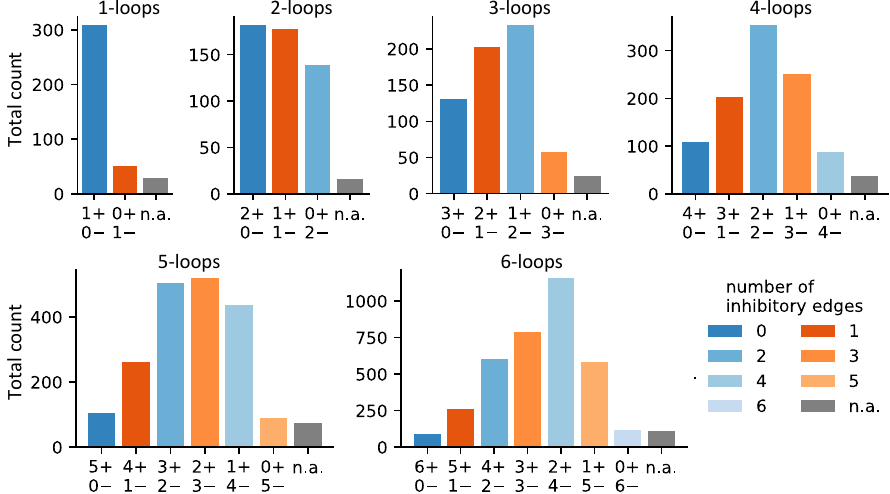}
\caption{{\bf Abundance of different types of feedback loops}. The total number of different types of FBLs is shown, stratified based on the number of involved genes (sub panels) and based on the number of activating ($+$) and inhibitory ($-$) regulations (x-axis). Loops, which contain conditional regulations preventing the determination of their type, are classified as n.a. Color indicates the number of negative edges in the FBL. Bars corresponding to positive FBLs (with an even number of negative interactions) are blue, while negative FBLs are red.}
\label{fig:feedback}
\end{figure}

Feedback loops (FBLs) constitute another important network motif. The parity of the number of inhibitory regulations determines if a FBL is \emph{positive} (even number) or \emph{negative} (uneven number). Each gene in a positive (negative) FBL exerts a positive (negative) effect on its own downstream expression. In general, negative FBLs buffer a perturbation and ensure homeostasis, while positive FBLs amplify perturbations and are necessary for bi- or multistationarity~\cite{thomas1990biological,thomas1995dynamical,kaern2005stochasticity}. We identified all FBLs involving up to six genes. For each FBL, we counted the number of activating and inhibitory regulations involved~(fig.~\ref{fig:feedback}). Just like FFLs, some FBLs contained conditional regulations, which prevented the determination of their exact type. As expected by chance, we found more complex loops than short 2-loops or even auto-regulatory loops (i.e., 1-loops). Also, FBLs with a balanced number of activating and inhibitory regulations are combinatorially more likely and were accordingly found more frequently. 


To compute an expected distribution for the number of activating versus inhibitory regulations in fixed-length FBLs, we employed two null models, which differ in the way that the proportion of activating regulations is computed. Null model 1 uses the same proportion for all FBLs within the same network, while null model 2 uses the fact that each FBL is contained in a strongly connected component (SCC) and derives the proportion of activating regulations only from this SCC (see Methods). 

\begin{figure} 
\centering
\includegraphics[width=3.55in]{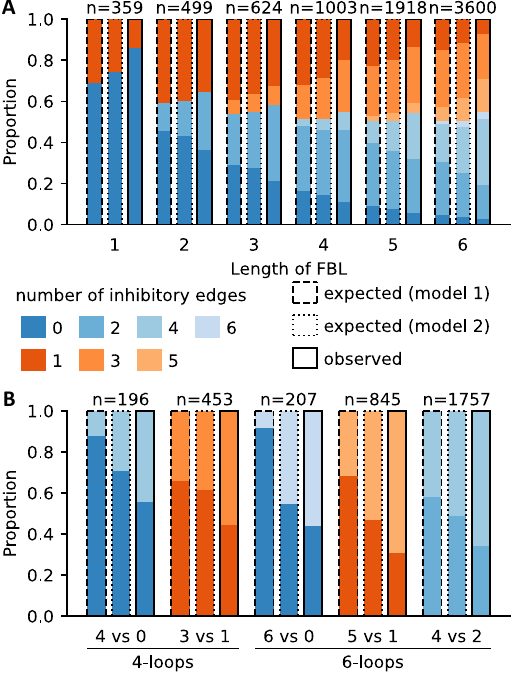}
\caption{{\bf Complex feedback loops are enriched for inhibitory edges}. (A) Stratification of all observed FBLs based on the number of involved genes (x-axis) and the number of activating versus inhibitory edges they contain (color). Positive FBLs are  blue, while negative FBLs are red. FBLs that contain conditional regulations are excluded. Each observed distribution (the rightmost of three bars with solid border) is compared to the expected distribution (left and middle bars with dashed and dotted borders), which is computed using two different null models (see Methods for details). n = total number of observed FBLs of a given length. (B) For 4- and 6-loops of the same type (positive or negative) and the same combinatorial likelihood, which depends on the number of activating versus inhibitory edges in the FBL, the observed relative abundance of FBLs with more activating versus more inhibitory edges is compared to the respective expected relative abundance, which is based on the same two null models as in A.}
\label{fig:fbl_negative_edges}
\end{figure}

\begin{figure}
    \centering
    \includegraphics[width=\textwidth]{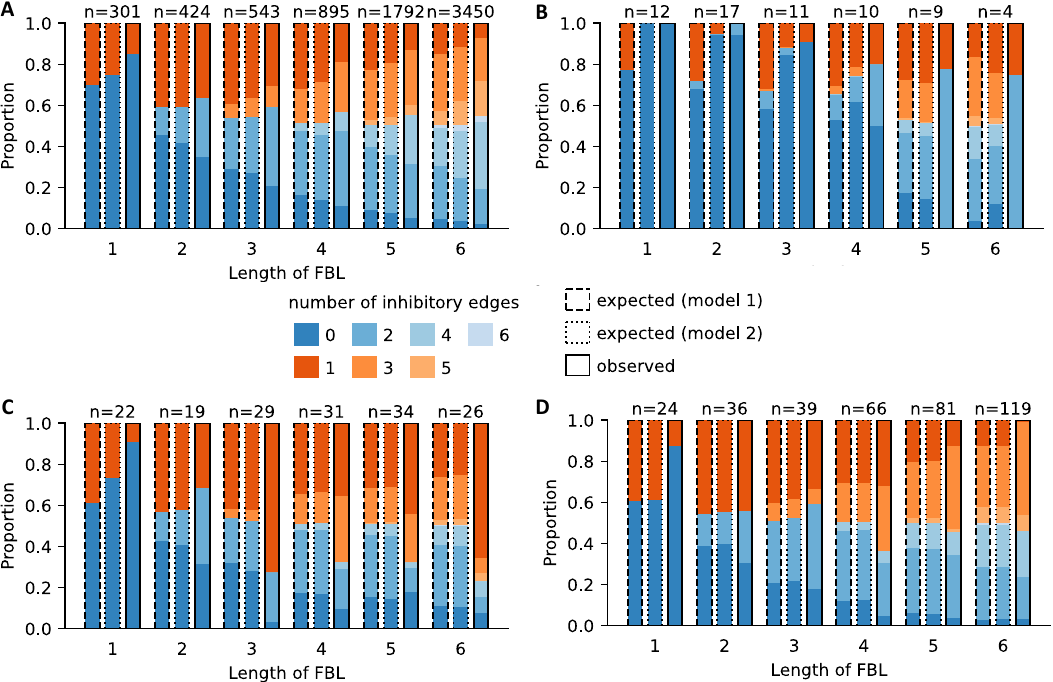}
    \caption{\new{{\bf Stratification of all observed feedback loops per kingdom.} All FBLs found in GRN models of (A) animals, (B) bacteria, (C) fungi, (D) plants are stratified based on the number of involved genes (x-axis) and the number of activating versus inhibitory edges they contain (color). Positive FBLs are  blue, while negative FBLs are red. FBLs that contain conditional regulations are excluded. Each observed distribution (the rightmost of three bars with solid border) is compared to the expected distribution (left and middle bars with dashed and dotted borders), which is computed using two different null models (see Methods for details). n = total number of observed FBLs of a given length. The corresponding inter-kingdom analysis is shown in Fig.~\ref{fig:fbl_negative_edges}A.}}
    \label{fig:kingdoms_fbls1}
\end{figure}

\begin{figure}
    \centering
    \includegraphics[width=\textwidth]{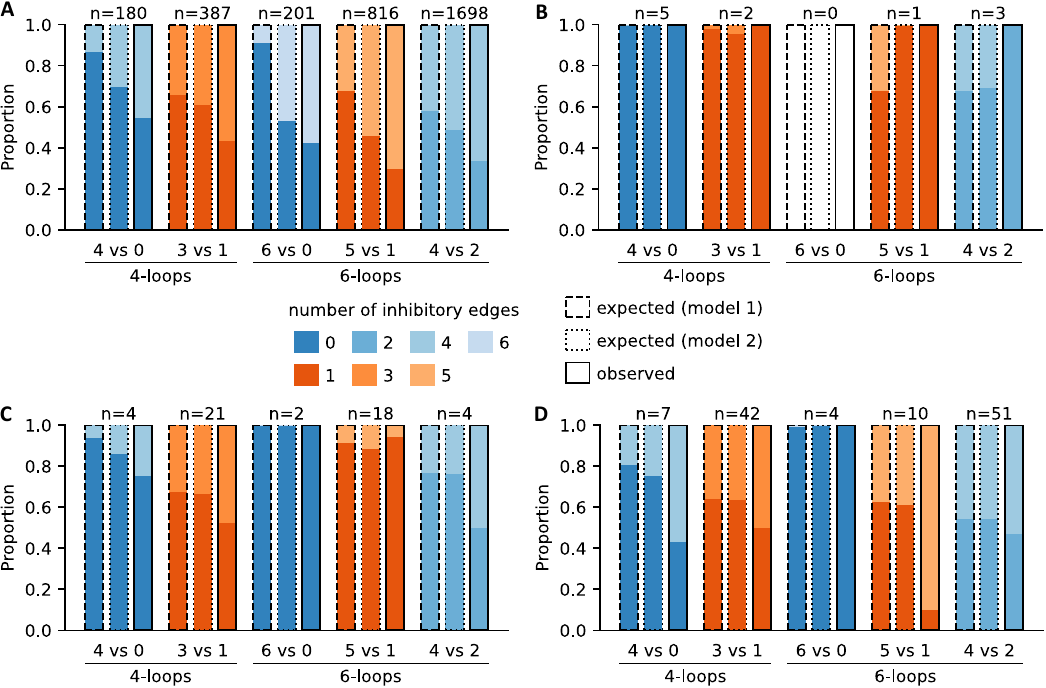}
    \caption{\new{{\bf Abundance of inhibitory edges in complex feedback loops per kingdom.} For FBLs of length 4 and 6 of the same type (positive or negative) and the same combinatorial likelihood, which depends on the number of activating versus inhibitory edges in the FBL, the observed relative abundance of FBLs with more activating versus more inhibitory edges is compared to the respective expected relative abundance. The expected distribution is computed using two different null models (see Methods for details). The analysis is stratified by kingdom: (A) animals, (B) bacteria, (C) fungi, (D) plants. The corresponding inter-kingdom analysis is shown in Fig.~\ref{fig:fbl_negative_edges}B.}}
    \label{fig:kingdoms_fbls2}
\end{figure}

For all different lengths, positive FBLs appeared slightly more frequently than expected~(Fig.~\ref{fig:fbl_negative_edges}A). We also observed more self-reinforcing than self-inhibitory regulations (1-loops) than expected. On the other hand, more complex FBLs containing two or more genes were enriched for inhibitory regulations. To enable an unbiased comparison, we considered specifically complex loops of the same type (positive or negative), with the same number of genes and the same number of combinatorially expected occurrences (that is, 4-loops with 4 versus none or 3 versus 1 inhibitory regulations, or 6-loops with 6 versus none, 5 versus 1 or 4 versus 2 inhibitory regulations). All five comparisons confirmed a surprising overabundance of negative regulations in the observed FBLs~(Fig.~\ref{fig:fbl_negative_edges}B). Notably, the differences between observed and expected relative abundances were consistently smaller (but still substantial) when considering null model 2. This aligns with our finding that most SCCs that contain many FBLs have a lower proportion of activating edges than the full network~(Supplementary Datasets 1 and \new[4]{5}). \new{Due to insufficient numbers of FBLs in non-animal GRN models, we were unable to assess the potential for kingdom-specific differences in the prevalence of specific types of FBLs~(figs.~\ref{fig:kingdoms_fbls1},\ref{fig:kingdoms_fbls2}).}

\subsection*{Criticality}
Gene regulation is a highly stochastic process due to e.g. low copy numbers of expressed molecules, random transitions between chromatin states, and extrinsic environmental perturbations~\cite{elowitz2002stochastic,kaern2005stochasticity}. \new[Despite these various sources of stochasticity, GRNs must maintain a stable phenotype in order to ensure consistent operation of the cellular processes.]{While some bacteria rely on noise in gene regulation to successfully mitigate risk through bet-hedging strategies~\cite{morawska2022diversity}, most GRNs are incentivized to maintain a stable phenotype in order to ensure consistent operation of the cellular processes, despite various sources of stochasticity.} At the same time, GRNs must be able to adapt to lasting changes in the environment. Due to this stability-evolvability tradeoff, GRNs have been hypothesized to operate in the so-called \emph{critical} dynamical regime, on the edge of order and chaos~\cite{aldana2007robustness}. Criticality has also been postulated for a variety of other biological networks such as neural networks or networks describing animal motion and social behavior~\cite{munoz2018colloquium}. The dynamical robustness of a Boolean network is typically measured by the average sensitivity or more general Derrida values~\cite{Derrida1,Derrida2}, which describe how a small perturbation affects the network over time. If, on average, the perturbation reduces in size after each gene has been synchronously updated once, the system operates in the \emph{ordered} regime; if it amplifies on average, the system is in the \emph{chaotic} regime, and if it remains, on average, of the similar size, the system exhibits \emph{criticality}. Many biological systems, modeled using Boolean networks, operate in the critical regime~\cite{balleza2008critical,daniels2018criticality}. 

\begin{figure*}
    \centering
\includegraphics[width=7.25in]{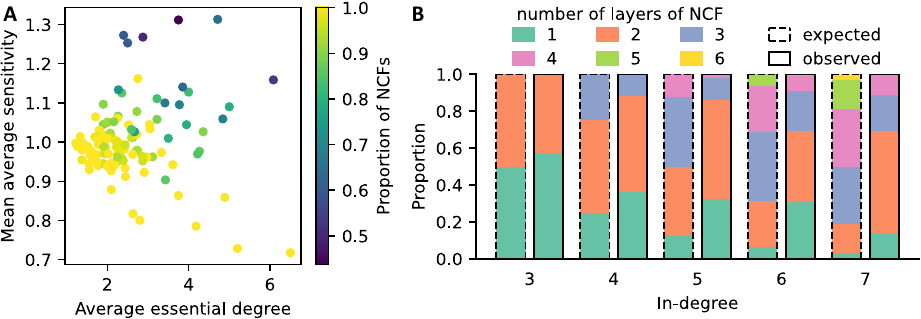}
\caption{{\bf Dynamical robustness of the GRN models}. (A) For each published model, the mean average sensitivity is plotted against the average number of essential regulators, colored by the proportion of model update rules that are NCFs. (B) Stratification of the observed NCFs by number of variables (x-axis) and layer structure (colored bars). The observed relative abundance (right bars with solid borders) is compared to the respective expected relative abundance (left bars with dotted borders).}
\label{fig:derrida}
\end{figure*}

For a synchronously updated Boolean network with $N$ nodes, the Derrida value for a single perturbation is simply the mean average sensitivity $s = 1/N \sum_{i=1}^N S(f_i)$ where $S(f_i)\in[0,n_i]$ is the average sensitivity of update function $f_i$ with $k_i$ inputs~\cite{Shmul04}. For random Boolean functions in $k$ (not necessarily essential) variables and with output bias $p$ (which describes the probability of activation, i.e., the probability of ones in the function's truth table), the expected average sensitivity is $2p(1-p)k$, and thus increases linearly in $k$. On the contrary, the expected average sensitivity of \new[NCFs]{nested canalizing functions (NCFs)} is $1$, irrespective of $k$~\cite{kadelka2017influence}. All 120 investigated models exhibited a mean average sensitivity near 1 (mean $= 1.0014$, standard deviation $= 0.09$), which constitutes the critical threshold between order and chaos~(Fig.~\ref{fig:derrida}A). 

\begin{figure} 
    \centering
\includegraphics[width=0.79\textwidth]{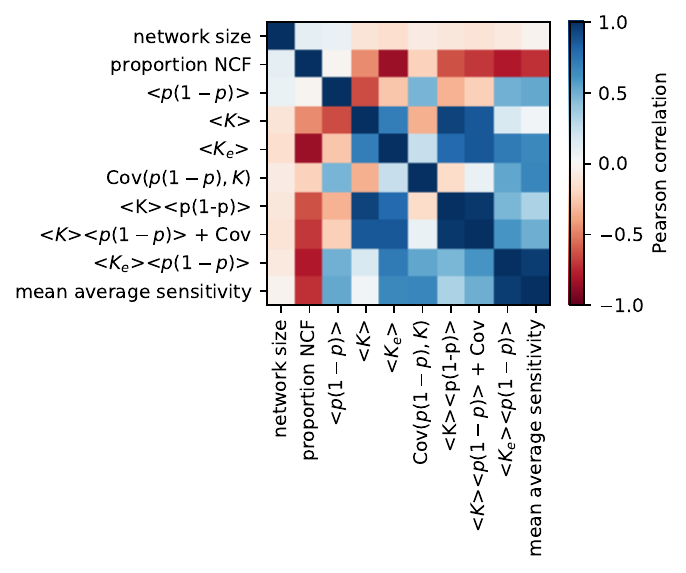}
\caption{{\bf Predictors of dynamical robustness}. Pairwise Pearson correlation between various properties and suggested predictors of dynamical robustness across the published GRN models. $<\cdot>$ denotes the mean, $p = $ output bias, $K = $ number of variables, \new{$K_e = $ effective connectivity, Cov $ = $ covariance of $p(1-p)$ and $K$}.}
\label{fig:derrida_correlations}
\end{figure}

Across the models, mean average sensitivity was not associated with average essential degree (Pearson's $r = 0.03$), nor with network size (Pearson's $r = -0.03$) but depended strongly on a model's proportion of update rules that were nested canalizing (Pearson's $r = -0.73$; fig.~\ref{fig:derrida_correlations}). The eight models with the lowest mean average sensitivity ($\leq 0.9$) were all completely governed by NCFs, while the five models with the highest mean average sensitivity ($\geq 1.17$) were among the models containing the lowest proportion of NCFs~(Fig.~\ref{fig:derrida}A).

\begin{figure} 
    \centering
\includegraphics[width=3.55in]{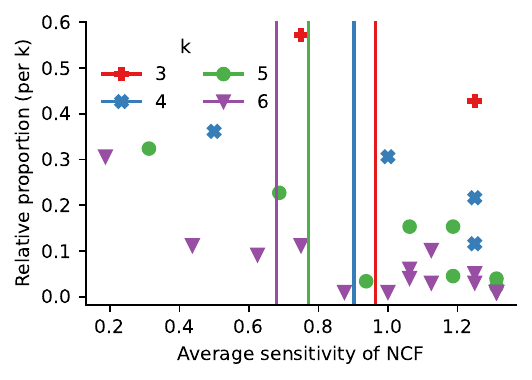}
\caption{{\bf Abundance of \new[different]{insensitive} NCFs}. The relative proportion of observed NCFs in $k=$3-6 variables, stratified by layer structure (exact numbers in table~\ref{tab:ncfs_observed}), is plotted against their average sensitivity (markers, with color differentiating $k$). For each $k$, the mean average sensitivity of all observed NCFs in $k$ variables is depicted by a vertical line.}
\label{fig:proportion_ncfs}
\end{figure}

This led us to investigate the relative frequency of different NCFs in the published models. Any \new[NCF (and more generally any]{} nonzero Boolean function\new[)]{} has a unique standard monomial form, in which all variables are distributed into canalizing layers of importance and a non-canalizing core~\cite{Yua1,he2016stratification}. 
\new{NCFs are specifically those Boolean functions where the core is empty, i.e., where all variables become eventually canalizing and possess a hierarchical importance order. To understand why NCFs appear frequently in GRNs, consider as an example a typical situation in gene regulation: two proteins X and Y can each independently initiate the transcription of a gene, as long as a repressor Z is not present to block the recruitment of RNA polymerase. The regulation of the gene in Boolean logic is best described by the NCF {(X OR Y) AND NOT Z}, which has two layers of importance, with $Z$ being most important.} 
NCFs with the same layer structure (i.e., with the same number of variables in each \new{canalizing} layer) have the same average sensitivity~\cite{kadelka2017influence, kadelka2017multistate}. For a given number of variables $k\geq 2$, there exists a bijection between $p(1-p)$ and the layer structure of an NCF, and there are $2^{k-2}$ NCFs with different layer structure, with each layer structure appearing equally likely by chance. Surprisingly, we \new[found]{discovered} a very non-equal occurrence among the NCFs in the published models~(table~\ref{tab:ncfs_observed}). Partially in line with the findings of high redundancy, NCFs with fewer layers appeared more frequently~(Fig.~\ref{fig:derrida}B). The observed NCFs also exhibited lower than expected mean average sensitivity~(Fig.~\ref{fig:proportion_ncfs}), and the higher the number of variables the lower was the observed mean average sensitivity. These findings suggest that biological networks are enriched for \new[nested canalizing functions]{NCFs} that induce stable dynamics as a means to counter-balance some less canalizing and more sensitive functions.
\new{While earlier studies suggested GRNs manage to operate in the critical regime due to the abundance of canalizing update rules~\cite{daniels2018criticality}, our results provide a more detailed understanding of this process, by pointing to NCFs with specific dynamic features as stabilizers of GRNs.}

For many years, an accurate description of the critical boundary in terms of macro- and micro-level network properties has received a lot of attention. Rather than considering a binary classification problem as in~\cite{manicka2022effective,costa2023effective}, we tested how well several suggested predictors of criticality correlated with the mean average sensitivity across this largest repository of published biological networks~(fig.~\ref{fig:derrida_correlations}). The first description of the critical boundary $2\langle k\rangle\langle p(1-p)\rangle = 1$~\cite{Shmul04}, where $\langle \cdot\rangle$ denotes the mean value across all rules within one model, only weakly correlated with the mean average sensitivity~(Pearson's r $= 0.31$). As described in~\cite{daniels2018criticality}, this is likely because it lacks to account for canalization, the essential in-degree and a negative correlation between $k$ and $p(1-p)$ in most models. Accounting for this covariance via $\langle k\rangle\langle p(1-p)\rangle + \text{Cov}$, as suggested in~\cite{daniels2018criticality}, led to a better correlation~(Pearson's r $= 0.49$). However, the covariance alone was even more correlated with the mean average sensitivity~(Pearson's r $= 0.66$). A predictor of the critical boundary that accounts for collective canalization by replacing $k$, the connectivity, with $K_e$, the effective connectivity, was recently suggested: $3.94\langle K_e\rangle\langle p(1-p)\rangle$~\cite{manicka2022effective}. This predictor correlated almost perfectly with the mean average sensitivity~(Pearson's r $= 0.95$), highlighting how well the effective connectivity captures the stabilizing effect of canalization on the dynamics of biological networks.

\section*{Conclusion}\label{sec-conclusion}
Gene expression constitutes the most fundamental process in which genotype determines phenotype. A detailed understanding of the design principles that regulate this process is therefore of great importance. We utilized combined knowledge from numerous experts in their respective fields to perform a meta-analysis of published gene regulatory networks. Boolean networks constituted the perfect modeling framework for this kind of analysis due to their simplicity, easy comparability, and widespread use. A large literature search yielded the most extensive database of expert-curated Boolean gene regulatory network models thus far, which may be queried to generate and test various types of hypotheses.

We highlighted the usefulness of this resource by focusing on several design principles of GRNs. We confirmed that the regulatory logic is not random but highly canalized. Using a broader definition of canalization, we showed that even regulatory interactions that were not considered canalizing in previous analyses, exhibited a high level of canalization. Canalization and genetic redundancy are two correlated concepts; GRNs proved to be independently enriched for both. We further studied the presence of small network motifs and discovered various types of motifs that were vastly more or less abundant than expected by chance. Finally, we provided strong evidence for the hypothesis that all GRNs operate dynamically close to the edge of order and chaos due to a tradeoff between stability and adaptability. \new{The abundance of nested canalizing update rules, specifically NCFs that are insensitive to perturbations, appeared to maintain critical dynamics for more densely connected GRNs.}

The described analysis suffers from several obvious limitations. First, not all biological phenomena can be accurately described in simple Boolean logic. There are a variety of published models that allow for more than two states. A similar analysis of more general models might provide more detailed insights into gene regulation but will itself suffer from the increased complexity of describing the studied concepts in the non-Boolean case. Second, there exists no feasible way to test the representativeness or completeness of our generated database of Boolean models. Even if a complete database of all published Boolean network models existed, the results would still be biased as some processes and species (e.g., model organisms) receive more attention and are modeled more frequently than others. Third, design principles of GRNs will likely differ among kingdoms of life or even among lower taxonomic levels. \new[A stratification of the results based on the organism under consideration would probably yield valuable information. We opted against such an analysis here as most (94, $77\%$) of the published models describe GRNs in animals.]{We therefore stratified the main analyses, wherever feasible, by kingdom. Since most of the published Boolean models and especially the large ones describe GRNs in animals, this meta-analysis lacks the statistical power to identify potential differences in design principles between kingdoms. In light of this, the identified design principles should primarily be understood as features of animal GRNs.}
\new{A last limitation lies in the study design itself. Since we analyze expert-curated Boolean GRN models, it is impossible to rule out the introduction of bias by the experts who built the models. Many of the trends and properties we identified are highly significant and consistent, which means they likely reflect true biological qualities of regulatory networks. However, to know for sure, future research is needed.}
\new{Since one of the main goals of synthetic biology is to generate complex networks with programmable functionality, synthetic biologists could, for example, engineer and study gene circuits that feature specific design principles suggested here. In addition, \textit{in silico} experiments could clarify if and how the suggested design principles are advantageous for GRNs.}

\section*{Methods}
\subsection*{Database creation}
Aiming to identify all published Boolean network models of GRNs, we developed an algorithm that parses all of the more than 30 million abstracts indexed in the literature search engine Pubmed and used keywords to rank the abstracts based on how likely they were to contain a Boolean network model. To identify the keywords, we relied on the Cell Collective, a pre-existing repository of Boolean network models, which, at the time of access, contained 78 Boolean network models published in 65 distinct papers~\cite{helikar2012cell}. The abstracts of these 65 papers served as a training set for the identification of keywords indicative of the presence of a Boolean network model. We considered as possible indicators (i) any word that occurred in at least two Cell Collective abstracts and was not among the most common 3000 words found in an English dictionary, (ii) all fixed combinations of two and three non-common words like ``logical modeling" or ``Boolean network model", and (iii) all co-occurrences of two or three single non-common words in the same abstract, e.g. the co-occurrence of the words ``logical", ``regulatory" and ``modelling" in an abstract, not necessarily in the same fixed order. While the use of an automatic British English to American English conversion tool may have helped to limit the number of indicators, we chose to treat words that are spelled differently in British and American English as two separate words. For any possible indicator, we calculated a quality score as the ratio of the number of Cell Collective abstracts in which it occurred over the total number of Pubmed abstracts containing this indicator. This procedure resulted in 1297 publications with at least one indicator with a quality score of $5\%$ or greater. We then manually investigated these 1297 publications to decide if they indeed contained a GRN model. During the manual review, an additional 369 referenced publications were investigated, as they were manually deemed to be of potential interest despite lacking an indicator with quality score $\geq 5\%$, resulting in a total of 1666 reviewed publications.

\subsubsection*{Model exclusion}
To avoid the introduction of various of kinds of bias into the analysis, we used the following strict criteria for the inclusion of models.
\begin{enumerate}
\item We excluded models where the update rules were solely generated using an inference method or where default updates like threshold rules were consistently used. Our goal was to include only models where the update rules were built based on biological expertise and knowledge gained from appropriate experiments.
\item In addition, identical models that were presented in multiple publications were only included once, and we aimed to include the earliest publication that initially presented the model. In total, 165 models passed this step and were extracted as described in the next subsection.
\item An automated quality check ensured that highly similar models were only included once in the analysis. The overlap index, also known as Szymkiewicz-Simpson coefficient, measures the overlap between two sets $A$ and $B$ and is defined as $|A \cap B|/\min(|A|,|B|)\in [0,1]$~\cite{szymkiewicz1934contribution}. We defined two models to be highly similar if the overlap between the set of their variables (with each variable expressed as a lower case string with '.', '-' and '\_' removed) was $\geq 90\%$. After single-linkage hierarchical clustering of highly similar models, we manually reviewed all clusters. For each cluster, we removed all but one model from the analysis, aiming to include the final version of the model in the analysis. Most frequently, this meant inclusion of the latest published model, or the last stated model for highly similar models stemming from the same publication. This additional quality control step led to the exclusion of 39 out of the 163 identified models.
\item Finally, we manually investigated the overlap between all models stemming from the same publication. For one publication, we removed two additional models as a third, included model from this publication was the combination of the two excluded models~\cite{thakar2012network}. Three other publications also contained more than one model. All these models were substantially different, as they described different GRNs or pathways with low overlap between the variables~\cite{mbodj2013logical,ryll2011large,der2014boolean}.
\end{enumerate}

\subsubsection*{Model extraction and standardization}
Boolean network models are presented in various formats in the literature. Using customized Python scripts, we extracted all published Boolean network models that were not excluded (see Model exclusion) and transformed them into a standardized format. In this format, each line describes the regulation of one gene; the name of the regulated gene is on the left, followed by ``=", followed by the Boolean update rule with operators AND, OR, and NOT. External parameters do not have an update rule and only occur in the update rules of the genes they regulate. For example,
\begin{align*}
\text{A} &= \text{B OR C}\\
\text{B} &= \text{A OR (C AND D)}\\
\text{C} &= \text{NOT A}
\end{align*}
represents a model with three genes, A, B and C, and one external parameter D.

\subsection*{Meta-analysis}
All analyses were performed in Python 3.10 using the libraries numpy, scipy, networkx, cana, matplotlib and itertools. In particular, we wrote a Python script, available at \href{https://github.com/ckadelka/DesignPrinciplesGeneNetworks}{https://github.com/ckadelka/DesignPrinciplesGeneNetworks}, which takes as input a Boolean model, described in standardized format, and returns, among other things, an adjacency matrix of the wiring diagram of the model, as well as completely evaluated update rules. That is, each update rule of $k$ inputs is represented as a vector of length $2^k$, which together with the wiring diagram enables all presented analyses. 

For computational reasons, we restricted most analyses to update rules with 20 or fewer inputs. The two models that each contained a single rule with more inputs (GLI1 in the hedgehog signaling pathway~\cite{chowdhury2013structural} is regulated by 24 inputs, while Shc in a multi-scale model of ErbB receptor signal transduction~\cite{helikar2013comprehensive} is even regulated by 27 inputs) were excluded from the network motif and criticality analyses, as the specific types of regulation (activation, inhibition, conditional) and number of essential inputs could not be determined for rules with so many inputs.

\subsection*{Measures of canalization}
This study includes several measures of canalization. By~\cite{kauffman1974large}, a Boolean function $f(x_1,\ldots,x_n): \{0,1\}^n \to \{0,1\}$ is canalizing if there exists a canalizing variable $x_i$, a canalizing input $a\in\{0,1\}$ and a canalized output $b\in\{0,1\}$ such that 
$$f(x_1,\ldots,x_n) = \begin{cases}b & \text{if } x_i = a,\\g(x_1,\ldots,x_{i-1},x_{i+1},\ldots,x_n)\not\equiv b & \text{otherwise.}\end{cases}$$
If the subfunction $g$ is also canalizing, then $f$ is $2$-canalizing, etc. More generally, $f$ is \emph{$k$-canalizing}, where $1 \leq k \leq n$, with respect to the permutation $\sigma \in \mathcal{S}_n$, inputs $a_1,\ldots,a_k$, and outputs $b_1,\ldots,b_k$ if
\begin{equation*}f(x_{1},\ldots,x_{n})=
\left\{\begin{array}[c]{ll}
b_{1} & x_{\sigma(1)} = a_1,\\
b_{2} & x_{\sigma(1)} \neq a_1, x_{\sigma(2)} = a_2,\\
b_{3} & x_{\sigma(1)} \neq a_1, x_{\sigma(2)} \neq a_2, x_{\sigma(3)} = a_3,\\
\vdots  & \vdots\\
b_{k} & x_{\sigma(1)} \neq a_1,\ldots,x_{\sigma(k-1)}\neq a_{k-1}, x_{\sigma(k)} = a_k,\\
f_C\not\equiv b_k & x_{\sigma(1)} \neq a_1,\ldots,x_{\sigma(k-1)}\neq a_{k-1}, x_{\sigma(k)} \neq a_k.
\end{array}\right.\end{equation*} 
Here, $f_C = f_C(x_{\sigma(k+1)},\ldots,x_{\sigma(n)})$ is the \emph{core function}, a Boolean function on $n-k$ variables. When $f_C$ is not canalizing, then the integer $k$ is the \emph{canalizing depth} of $f$~\cite{layne2012nested}. If $k=n$ (i.e., if all variables are become eventually canalizing), then $f$ is a \emph{nested canalizing function} (NCF)~\cite{Kau2}.
By~\cite{he2016stratification}, every nonzero Boolean function $f(x_1,\ldots,x_n)$ can be uniquely written as 
\begin{equation*}
      f(x_1,\ldots,x_n) = M_1(M_2(\cdots (M_{r-1}(M_rp_C + 1) + 1)\cdots)+ 1)+ q,  
\end{equation*}
where each $M_i = \prod_{j=1}^{k_i} (x_{i_j} + a_{i_j})$ is a non-constant extended monomial, $p_C$ is the \emph{core polynomial} of $f$, and $k = \sum_{i=1}^r k_i$ is the canalizing depth. Each $x_i$ appears in exactly one of $\{M_1,\ldots,M_r,p_C\}$. The \emph{layer structure} of $f$ is the vector $(k_1,k_2,\ldots,k_r)$ and describes the number of variables in each layer $M_i$~\cite{kadelka2017influence, dimitrova2022revealing}.

More recently, canalization has been considered as a property of the Boolean function, rather than on the variable level~\cite{Reichhardt}. In~\cite{gates2021effective}, canalization is equated to input redundancy, enabling the definition of variable/edge- and function/node-level properties, used in this study, such as the \emph{edge effectiveness} and the \emph{effective connectivity}. The \emph{canalizing strength} constitutes an alternative approach to measure canalization on the function level~\cite{kadelka2023collectively}. This approach generalizes Kauffman's original definition of canalization more closely. For brevity, we refer the interested reader to these papers for details. 

\subsection*{Expected number of loops}
The likelihood of a specific FFL or FBL type depends on the ratio of positive versus negative edges. Due to substantial variation of this ratio across models (Supplementary Dataset 1), we computed the expected distribution of specific FFL and FBL types separately for each model. For model $i$, let $p_i\in [0,1]$ denote the proportion of activating edges (out of all activating and inhibitory edges, excluding conditional and non-essential edges). 

To compute the expected number of different FFLs in model $i$, let $n_i$ and $n^{t}_i$ denote the total number of FFLs and the total number of FFLs of type $t$, respectively. To create a null expectation, we assume that each edge is activating with probability $p_i$ and inhibitory with probability $1-p_i$. Then,
$$\mathbb{E}\big[n^{t}_i \big| n_i\big] = n_i p_i^{a(t)} (1-p_i)^{3-a(t)},$$
where $a(t)\in\{0,1,2,3\}$ denotes the number of activating edges in FFLs of type $t$. The expected number of FFLs of type $t$ across all models is simply the sum of all model-specific expected numbers. This is null model 1.

Null model 1 can also be used to compute the expected number of different FBLs. Let $n^k_i$ and $n^{k,j}_i$ denote the total number of $k$-loops and the total number of $k$-loops containing exactly $j$ inhibitory edges, respectively. Then,
$$\mathbb{E}\big[n^{k,j}_i \big| n^k_i\big] = n^k_i \binom{k}{j} p_i^{k-j} (1-p_i)^{j}.$$
The expected number of $k$-loops containing exactly $j$ inhibitory edges across all models is the sum of all model-specific expected numbers.

Null model 2 differs in the way the proportion of activating edges is computed. It uses the fact that each FBL is part of a strongly connected component (SCC). Rather than using one overall proportion per model, null model 2 bases the expectation on the proportion within each FBL's SCC. Let $p_{i,c}\in[0,1]$ denote the proportion of activating edges in SCC $c$ (out of all activating and inhibitory edges, excluding conditional and non-essential edges). Let $\ell_r, r=1,\ldots,n^k_i$ denote all $k$-loops of model $i$ and let $c(\ell_r)$ denote the strongly connected component containing $\ell_r$. Then,
$$\mathbb{E}\big[n^{k,j}_i \big| n^k_i\big] = \sum_{k\text{-loops}\ \ell_r} \binom{k}{j} p_{i,c(\ell_r)}^{k-j} (1-p_{i,c(\ell_r)})^{j}.$$
As before, the expected number of $k$-loops containing exactly $j$ inhibitory edges across all models is the sum of all model-specific expected numbers.

\subsection*{Dynamical robustness}
As an indicator of the dynamical robustness of a Boolean network $F$, we computed the mean average sensitivity $s$, which describes the average size of an initial perturbation of size 1 after each gene has been synchronously updated once. That is,
$$s = \mathbb{E}\big[d(F(\mathbf x), F(\mathbf y))\big| d(\mathbf x,\mathbf y) = 1\big],$$
where $d$ is the Hamming distance between two binary states. For nested canalizing networks, there exists an exact formula for $s$~\cite{kadelka2017influence}. For all biological networks that were not entirely governed by NCFs, we relied instead on simulations to estimate $s$. For each network, we generated 10,000 random states $\mathbf x \in \{0,1\}^{N+E}$ where $N$ is the number of genes and $E$ the number of external parameters. For each state, we selected a random gene $i\in\{1,2,\ldots,N\}$ to be flipped to generate $\mathbf y = \mathbf x + \mathbf{e_i}$ with $d(\mathbf x,\mathbf y) = 1$.

\bibliography{ncf_bib_latest}

\begin{thebibliography}{10}

\bibitem{stelling2004robustness}
J.~Stelling, U.~Sauer, Z.~Szallasi, F.~J. Doyle~III, J.~Doyle, Robustness of
  cellular functions, {\it Cell\/} {\bf 118}, 675--685 (2004).

\bibitem{shen2002network}
S.~S. Shen-Orr, R.~Milo, S.~Mangan, U.~Alon, Network motifs in the
  transcriptional regulation network of escherichia coli, {\it Nature
  genetics\/} {\bf 31}, 64--68 (2002).

\bibitem{milo2002network}
R.~Milo, {\it et~al.\/}, Network motifs: simple building blocks of complex
  networks, {\it Science\/} {\bf 298}, 824--827 (2002).

\bibitem{alon2007network}
U.~Alon, Network motifs: theory and experimental approaches, {\it Nature
  Reviews Genetics\/} {\bf 8}, 450--461 (2007).

\bibitem{gerstein2012architecture}
M.~B. Gerstein, {\it et~al.\/}, Architecture of the human regulatory network
  derived from encode data, {\it Nature\/} {\bf 489}, 91--100 (2012).

\bibitem{mangan2003structure}
S.~Mangan, U.~Alon, Structure and function of the feed-forward loop network
  motif, {\it Proceedings of the National Academy of Sciences\/} {\bf 100},
  11980--11985 (2003).

\bibitem{cotterell2010atlas}
J.~Cotterell, J.~Sharpe, An atlas of gene regulatory networks reveals multiple
  three-gene mechanisms for interpreting morphogen gradients, {\it Molecular
  systems biology\/} {\bf 6}, 425 (2010).

\bibitem{mangan2003coherent}
S.~Mangan, A.~Zaslaver, U.~Alon, The coherent feedforward loop serves as a
  sign-sensitive delay element in transcription networks, {\it Journal of
  molecular biology\/} {\bf 334}, 197--204 (2003).

\bibitem{nowak1997evolution}
M.~A. Nowak, M.~C. Boerlijst, J.~Cooke, J.~M. Smith, Evolution of genetic
  redundancy, {\it Nature\/} {\bf 388}, 167--171 (1997).

\bibitem{gu2003role}
Z.~Gu, {\it et~al.\/}, Role of duplicate genes in genetic robustness against
  null mutations, {\it Nature\/} {\bf 421}, 63--66 (2003).

\bibitem{kauffman1969metabolic}
S.~A. Kauffman, Metabolic stability and epigenesis in randomly constructed
  genetic nets, {\it Journal of theoretical biology\/} {\bf 22}, 437--467
  (1969).

\bibitem{daniels2018criticality}
B.~C. Daniels, {\it et~al.\/}, Criticality distinguishes the ensemble of
  biological regulatory networks, {\it Physical review letters\/} {\bf 121},
  138102 (2018).

\bibitem{Wad}
C.~H. Waddington, Canalization of development and the inheritance of acquired
  characters, {\it Nature\/} {\bf 150}, 563--565 (1942).

\bibitem{schwab2020concepts}
J.~D. Schwab, S.~D. K{\"u}hlwein, N.~Ikonomi, M.~K{\"u}hl, H.~A. Kestler,
  Concepts in boolean network modeling: What do they all mean?, {\it
  Computational and Structural Biotechnology Journal\/}  (2020).

\bibitem{karlebach2008modelling}
G.~Karlebach, R.~Shamir, Modelling and analysis of gene regulatory networks,
  {\it Nature Reviews Molecular Cell Biology\/} {\bf 9}, 770--780 (2008).

\bibitem{matys2003transfac}
V.~Matys, {\it et~al.\/}, {TRANSFAC}: transcriptional regulation, from patterns
  to profiles, {\it Nucleic acids research\/} {\bf 31}, 374--378 (2003).

\bibitem{khan2018jaspar}
A.~Khan, {\it et~al.\/}, {JASPAR} 2018: update of the open-access database of
  transcription factor binding profiles and its web framework, {\it Nucleic
  acids research\/} {\bf 46}, D260--D266 (2018).

\bibitem{gama2016regulondb}
S.~Gama-Castro, {\it et~al.\/}, Regulon{DB} version 9.0: high-level integration
  of gene regulation, coexpression, motif clustering and beyond, {\it Nucleic
  acids research\/} {\bf 44}, D133--D143 (2016).

\bibitem{helikar2012cell}
T.~Helikar, {\it et~al.\/}, The cell collective: toward an open and
  collaborative approach to systems biology, {\it BMC systems biology\/} {\bf
  6}, 96 (2012).

\bibitem{manicka2023nonlinearity}
S.~Manicka, K.~Johnson, M.~Levin, D.~Murrugarra, The nonlinearity of regulation
  in biological networks, {\it NPJ Systems Biology and Applications\/} {\bf 9},
  10 (2023).

\bibitem{gates2021effective}
A.~J. Gates, R.~Brattig~Correia, X.~Wang, L.~M. Rocha, The effective graph
  reveals redundancy, canalization, and control pathways in biochemical
  regulation and signaling, {\it Proceedings of the National Academy of
  Sciences\/} {\bf 118}, e2022598118 (2021).

\bibitem{subbaroyan2022minimum}
A.~Subbaroyan, O.~C. Martin, A.~Samal, Minimum complexity drives regulatory
  logic in boolean models of living systems, {\it PNAS nexus\/} {\bf 1},
  pgac017 (2022).

\bibitem{manicka2022effective}
S.~Manicka, M.~Marques-Pita, L.~M. Rocha, Effective connectivity determines the
  critical dynamics of biochemical networks, {\it Journal of the Royal Society
  Interface\/} {\bf 19}, 20210659 (2022).

\bibitem{costa2023effective}
F.~X. Costa, J.~C. Rozum, A.~M. Marcus, L.~M. Rocha, Effective connectivity and
  bias entropy improve prediction of dynamical regime in automata networks,
  {\it Entropy\/} {\bf 25}, 374 (2023).

\bibitem{albert2002statistical}
R.~Albert, A.-L. Barab{\'a}si, Statistical mechanics of complex networks, {\it
  Reviews of modern physics\/} {\bf 74}, 47 (2002).

\bibitem{barabasi1999emergence}
A.-L. Barab{\'a}si, R.~Albert, Emergence of scaling in random networks, {\it
  Science\/} {\bf 286}, 509--512 (1999).

\bibitem{guelzim2002topological}
N.~Guelzim, S.~Bottani, P.~Bourgine, F.~K{\'e}p{\`e}s, Topological and causal
  structure of the yeast transcriptional regulatory network, {\it Nature
  genetics\/} {\bf 31}, 60--63 (2002).

\bibitem{luscombe2004genomic}
N.~M. Luscombe, {\it et~al.\/}, Genomic analysis of regulatory network dynamics
  reveals large topological changes, {\it Nature\/} {\bf 431}, 308--312 (2004).

\bibitem{bemer2017cross}
M.~Bemer, A.~D. van Dijk, R.~G. Immink, G.~C. Angenent, Cross-family
  transcription factor interactions: an additional layer of gene regulation,
  {\it Trends in plant science\/} {\bf 22}, 66--80 (2017).

\bibitem{raeymaekers2002dynamics}
L.~Raeymaekers, Dynamics of boolean networks controlled by biologically
  meaningful functions, {\it Journal of Theoretical Biology\/} {\bf 218},
  331--341 (2002).

\bibitem{struhl1999fundamentally}
K.~Struhl, Fundamentally different logic of gene regulation in eukaryotes and
  prokaryotes, {\it Cell\/} {\bf 98}, 1--4 (1999).

\bibitem{gibson2000canalization}
G.~Gibson, G.~Wagner, Canalization in evolutionary genetics: a stabilizing
  theory?, {\it BioEssays\/} {\bf 22}, 372--380 (2000).

\bibitem{hallgrimsson2019developmental}
B.~Hallgrimsson, {\it et~al.\/}, {\it Seminars in cell \& developmental
  biology\/} (Elsevier, 2019), vol.~88, pp. 67--79.

\bibitem{kauffman1974large}
S.~Kauffman, The large scale structure and dynamics of gene control circuits:
  an ensemble approach, {\it Journal of Theoretical Biology\/} {\bf 44},
  167--190 (1974).

\bibitem{he2016stratification}
Q.~He, M.~Macauley, Stratification and enumeration of {B}oolean functions by
  canalizing depth, {\it Physica D: Nonlinear Phenomena\/} {\bf 314}, 1--8
  (2016).

\bibitem{kadelka2017influence}
C.~Kadelka, J.~Kuipers, R.~Laubenbacher, The influence of canalization on the
  robustness of {B}oolean networks, {\it Physica D: Nonlinear Phenomena\/} {\bf
  353}, 39--47 (2017).

\bibitem{harris2002model}
S.~E. Harris, B.~K. Sawhill, A.~Wuensche, S.~Kauffman, A model of
  transcriptional regulatory networks based on biases in the observed
  regulation rules, {\it Complexity\/} {\bf 7}, 23--40 (2002).

\bibitem{layne2012nested}
L.~Layne, E.~Dimitrova, M.~Macauley, Nested canalyzing depth and network
  stability, {\it Bulletin of mathematical biology\/} {\bf 74}, 422--433
  (2012).

\bibitem{dimitrova2022revealing}
E.~Dimitrova, B.~Stigler, C.~Kadelka, D.~Murrugarra, Revealing the canalizing
  structure of boolean functions: Algorithms and applications, {\it
  Automatica\/} {\bf 146}, 110630 (2022).

\bibitem{Reichhardt}
C.~O. Reichhardt, K.~E. Bassler, Canalization and symmetry in {B}oolean models
  for genetic regulatory networks, {\it Journal of Physics A: Mathematical and
  Theoretical\/} {\bf 40}, 4339 (2007).

\bibitem{kadelka2023collectively}
C.~Kadelka, B.~Keilty, R.~Laubenbacher, Collectively canalizing {B}oolean
  functions, {\it Advances in Applied Mathematics\/} {\bf 145}, 102475 (2023).

\bibitem{kaplan2008incoherent}
S.~Kaplan, A.~Bren, E.~Dekel, U.~Alon, The incoherent feed-forward loop can
  generate non-monotonic input functions for genes, {\it Molecular systems
  biology\/} {\bf 4}, 203 (2008).

\bibitem{gorochowski2018organization}
T.~E. Gorochowski, C.~S. Grierson, M.~di~Bernardo, Organization of feed-forward
  loop motifs reveals architectural principles in natural and engineered
  networks, {\it Science advances\/} {\bf 4}, eaap9751 (2018).

\bibitem{thomas1990biological}
R.~Thomas, R.~d'Ari, {\it Biological feedback\/} (CRC press, 1990).

\bibitem{thomas1995dynamical}
R.~Thomas, D.~Thieffry, M.~Kaufman, Dynamical behaviour of biological
  regulatory networks?i. biological role of feedback loops and practical use of
  the concept of the loop-characteristic state, {\it Bulletin of mathematical
  biology\/} {\bf 57}, 247--276 (1995).

\bibitem{kaern2005stochasticity}
M.~Kaern, T.~C. Elston, W.~J. Blake, J.~J. Collins, Stochasticity in gene
  expression: from theories to phenotypes, {\it Nature Reviews Genetics\/} {\bf
  6}, 451--464 (2005).

\bibitem{elowitz2002stochastic}
M.~B. Elowitz, A.~J. Levine, E.~D. Siggia, P.~S. Swain, Stochastic gene
  expression in a single cell, {\it Science\/} {\bf 297}, 1183--1186 (2002).

\bibitem{morawska2022diversity}
L.~P. Morawska, J.~A. Hernandez-Valdes, O.~P. Kuipers, Diversity of bet-hedging
  strategies in microbial communities—recent cases and insights, {\it WIREs
  Mechanisms of Disease\/} {\bf 14}, e1544 (2022).

\bibitem{aldana2007robustness}
M.~Aldana, E.~Balleza, S.~Kauffman, O.~Resendiz, Robustness and evolvability in
  genetic regulatory networks, {\it Journal of theoretical biology\/} {\bf
  245}, 433--448 (2007).

\bibitem{munoz2018colloquium}
M.~A. Munoz, Colloquium: Criticality and dynamical scaling in living systems,
  {\it Reviews of Modern Physics\/} {\bf 90}, 031001 (2018).

\bibitem{Derrida1}
B.~Derrida, G.~Weisbuch, Evolution of overlaps between configurations in random
  {B}oolean networks, {\it Journal de Physique\/} {\bf 47}, 1297--1303 (1986).

\bibitem{Derrida2}
B.~Derrida, Y.~Pomeau, Random networks of automata: a simple annealed
  approximation, {\it Europhysics Letters\/} {\bf 1}, 45 (1986).

\bibitem{balleza2008critical}
E.~Balleza, {\it et~al.\/}, Critical dynamics in genetic regulatory networks:
  examples from four kingdoms, {\it PLoS One\/} {\bf 3}, e2456 (2008).

\bibitem{Shmul04}
I.~Shmulevich, S.~A. Kauffman, Activities and sensitivities in {B}oolean
  network models, {\it Physical Review Letters\/} {\bf 93}, 048701 (2004).

\bibitem{Yua1}
Y.~Li, J.~O. Adeyeye, D.~Murrugarra, B.~Aguilar, R.~Laubenbacher, Boolean
  nested canalizing functions: A comprehensive analysis, {\it Theoretical
  Computer Science\/} {\bf 481}, 24--36 (2013).

\bibitem{kadelka2017multistate}
C.~Kadelka, Y.~Li, J.~Kuipers, J.~O. Adeyeye, R.~Laubenbacher, Multistate
  nested canalizing functions and their networks, {\it Theoretical Computer
  Science\/} {\bf 675}, 1--14 (2017).

\bibitem{szymkiewicz1934contribution}
D.~Szymkiewicz, Une contribution statistique {\`a} la g{\'e}ographie
  floristique, {\it Acta Societatis Botanicorum Poloniae\/} {\bf 11}, 249--265
  (1934).

\bibitem{thakar2012network}
J.~Thakar, A.~K. Pathak, L.~Murphy, R.~Albert, I.~M. Cattadori, Network model
  of immune responses reveals key effectors to single and co-infection dynamics
  by a respiratory bacterium and a gastrointestinal helminth, {\it PLoS
  computational biology\/} {\bf 8}, e1002345 (2012).

\bibitem{mbodj2013logical}
A.~Mbodj, G.~Junion, C.~Brun, E.~E. Furlong, D.~Thieffry, Logical modelling of
  {D}rosophila signalling pathways, {\it Molecular biosystems\/} {\bf 9},
  2248--2258 (2013).

\bibitem{ryll2011large}
A.~Ryll, R.~Samaga, F.~Schaper, L.~G. Alexopoulos, S.~Klamt, Large-scale
  network models of {IL}-1 and {IL}-6 signalling and their hepatocellular
  specification, {\it Molecular biosystems\/} {\bf 7}, 3253--3270 (2011).

\bibitem{der2014boolean}
S.~V. der Heyde, {\it et~al.\/}, Boolean {ErbB} network reconstructions and
  perturbation simulations reveal individual drug response in different breast
  cancer cell lines, {\it BMC systems biology\/} {\bf 8}, 1--22 (2014).

\bibitem{chowdhury2013structural}
S.~Chowdhury, R.~N. Pradhan, R.~R. Sarkar, Structural and logical analysis of a
  comprehensive hedgehog signaling pathway to identify alternative drug targets
  for glioma, colon and pancreatic cancer, {\it PLoS One\/} {\bf 8}, e69132
  (2013).

\bibitem{helikar2013comprehensive}
T.~Helikar, {\it et~al.\/}, A comprehensive, multi-scale dynamical model of
  {ErbB} receptor signal transduction in human mammary epithelial cells, {\it
  PLoS One\/} {\bf 8}, e61757 (2013).

\bibitem{Kau2}
S.~Kauffman, C.~Peterson, B.~Samuelsson, C.~Troein, Random {B}oolean network
  models and the yeast transcriptional network, {\it Proceedings of the
  National Academy of Sciences\/} {\bf 100}, 14796--14799 (2003).

\end{thebibliography}
\bibliographystyle{Science}

\subsection*{Acknowledgments}

We thank Audrey McCombs for helpful comments on an initial version of the manuscript\new[.]{, and Reinhard Laubenbacher for several helpful discussions.}
\newline

\new{
\noindent {\bf Funding:} Apart from travel support from the Simons Foundation to C.K. (grant number 712537), the authors acknowledge that they received no funding in support for this research.
\newline

\noindent {\bf Author Contributions:} Conceptualization: CK; Methodology: CK; Software: CK and AS; Investigation: CK, TMB, EH, JK, AS, and HS; Visualization: CK; Writing - Original Draft: CK; Writing - Review \& Editing: CK
\newline

\noindent {\bf Competing interests:} The authors declare that they have no competing interests.
}
\newline

\new{
\noindent {\bf Data and Materials Availability:} The database of all 163 extracted Boolean GRN models (in one standardized format), as well as Python source code and instructions for their analysis can be found at \href{https://github.com/ckadelka/DesignPrinciplesGeneNetworks}{github.com/ckadelka/DesignPrinciplesGeneNetworks}, which is archived at \href{https://zenodo.org/record/8310222}{zenodo.org/record/8310222}. 

Moreover, \href{https://booleangenenetworks.math.iastate.edu}{https://booleangenenetworks.math.iastate.edu} features an interactive website that enables users without programming experience to analyze all models (or a subset thereof).} Full bibliographical information for all 122 models included in this meta-analysis is available in Supplementary Dataset 1.

\section*{Supplementary Material}

\noindent {\bf This PDF file includes:}

\noindent Figs. S1 to S11

\noindent Tables S1 and S2

\clearpage

\section*{Supplementary Material}

\setcounter{page}{1}

\beginsupplement

\clearpage

\clearpage

\clearpage

\clearpage

\clearpage

\clearpage

\clearpage

\clearpage

\clearpage

\clearpage

\clearpage

\begin{table} 
    \centering
\caption{{\bf High prevalence of canalization (full table)}. Stratification of all identified update rules based on the number of essential inputs (rows) and the canalizing depth (columns). The color gradient is computed separately for each row.}
\includegraphics[width=\textwidth]{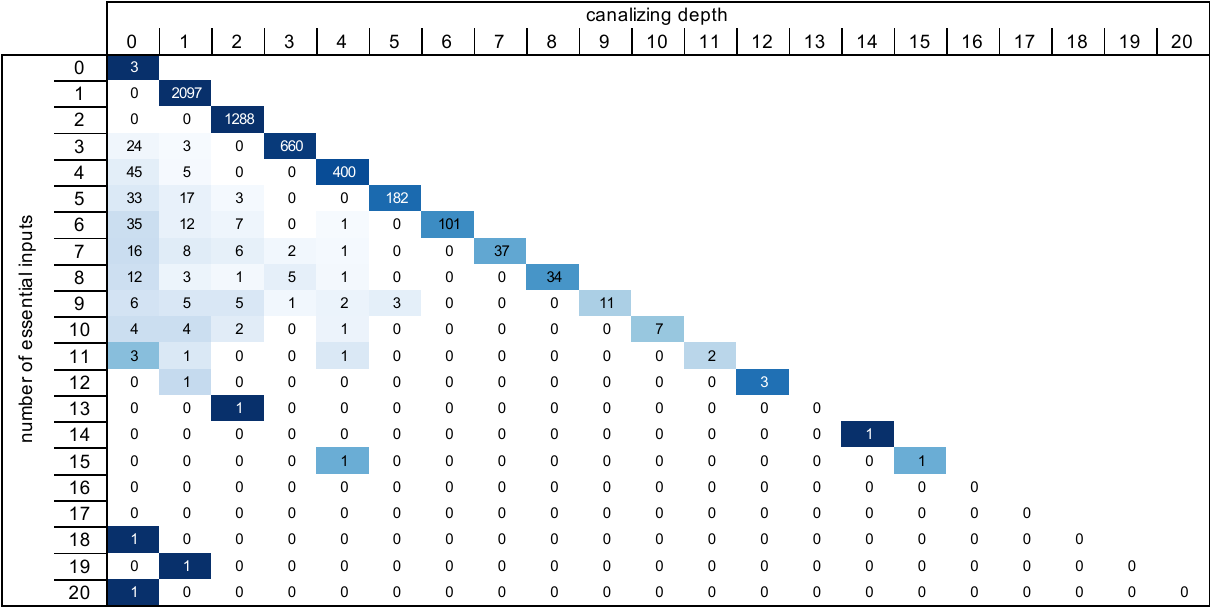}
\label{tab:canalization_full}
\end{table}

\clearpage

\begin{table} 
    \centering
\caption{{\bf Observed number of NCFs with 3-6 inputs, stratified by layer structure.} \new{The number of NCFs with a given number of inputs (3-6) and a given canalizing layer structure is shown. NCFs with the same layer structure have the same dynamical properties. All layer structures appear equally likely by chance. Green (orange) count data indicates more (fewer) than expected NCFs with a given layer structure. $K_e$ describes the effective connectivity and $p$ the output bias of the NCF.} }
\includegraphics[width=0.8\textwidth]{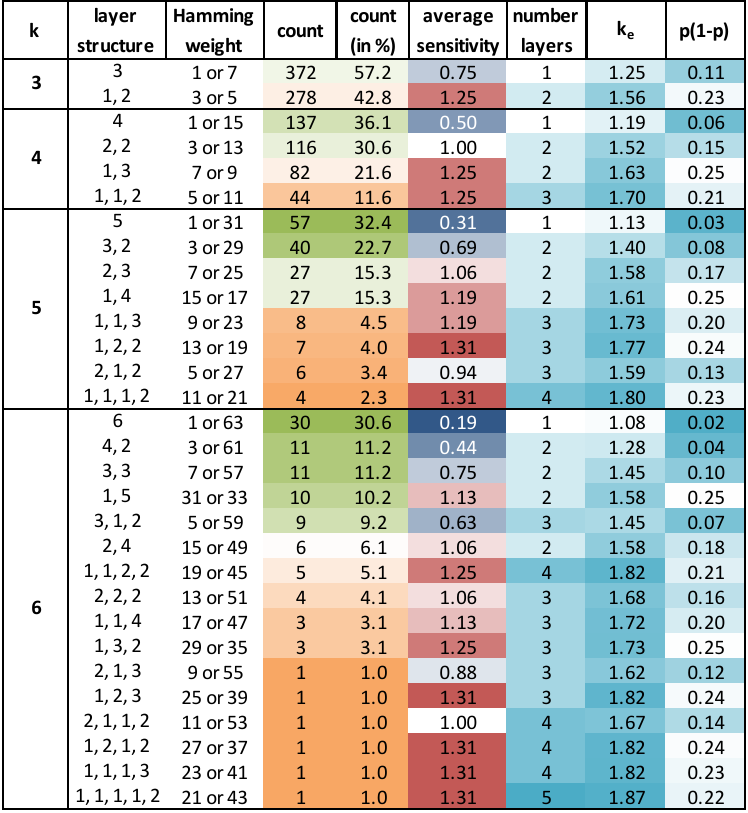}
\label{tab:ncfs_observed}
\end{table}

\clearpage

\noindent {\bf Supplementary dataset 1} General information about the 122 models included in the meta-analysis.

\noindent {\bf Supplementary dataset 2} Number of occurrences of specific nodes in the 122 models included in the meta-analysis.

\noindent {\bf Supplementary dataset 3} Abundance of different types of feed-forward loops in the 122 models included in the meta-analysis.

\noindent {\bf Supplementary dataset 4} Abundance of different types of clusters of feed-forward loops in the 122 models included in the meta-analysis.

\noindent {\bf Supplementary dataset 5} Abundance of different types of feedback loops in the 122 models included in the meta-analysis.

\end{document}